\documentclass{emulateapj}
\begin{document}
\title{New Estimators of Black Hole Mass in Active Galactic Nuclei with Hydrogen Paschen Lines}

\author{\textbf{Dohyeong Kim}\altaffilmark{1}, \textbf{Myungshin Im}\altaffilmark{1}
 and \textbf{Minjin Kim}\altaffilmark{2}} 

\altaffiltext{1}{Center for the Exploration of the Origin of the Universe (CEOU), 
Astronomy Program, Department of Physics and Astronomy, Seoul National University, 
Shillim-Dong, Kwanak-Gu, Seoul 151-742, South Korea}
\altaffiltext{2}{National Radio Astronomy Observatory, 520 Edgemont Road, 
Charlottesville, VA 22903, USA}

\begin{abstract}
 More than 50$\%$ of Active Galactic Nuclei (AGNs) are suspected to be red and affected by
 dust-obscuration.
 Meanwhile, popular spectral diagnostics of AGNs are based on optical or ultraviolet
 light, making the dust obscuration as a primary concern for understanding
 the general nature of AGNs and supermassive black holes residing in them.
 To provide with a method of investigating properties of the dusty AGNs,
 we derive new black hole (BH) mass estimators based on velocity widths
 and luminosities of Near Infrared (NIR) hydrogen emission lines 
 such as P$\alpha$ and P$\beta$, and also investigate
 the line ratios of these Hydrogen lines.
  To derive the BH mass ($M_{\rm BH}$) estimators, we used a sample of 37 
 unobscured Type-1 AGNs with a $M_{\rm BH}$ range of $10^{6.8}$--$10^{9.4}~M_{\odot}$,
 where $M_{\rm BH}$ come from either reverberation mapping method or single-epoch measurement 
 method using Balmer lines. Our work shows that $M_{\rm BH}$ can be estimated from
 the Paschen line luminosities and the velocity widths to the accuracy of 
 0.18 - 0.24 dex (rms scatter).
  We also show that the mean line ratios of the Paschen
 lines and the Balmer lines are $\mathrm{\frac{H\alpha}{P\alpha}} \simeq 9.00$,
 $\mathrm{\frac{H\beta}{P\alpha}} \simeq 2.70$, which are consistent with a Case B recombination
 under a typical AGN broad line region environment. These ratios can be used as reference 
 points when estimating the amount of dust extinction over the broad line region (BLR)
 for red AGNs.
 We expect the future application of the new BH mass estimators on red, dusty AGNs
 to provide a fresh view of obscured AGNs. 

\end{abstract}

\section{INTRODUCTION}

 Supermassive black holes (SMBHs), found at centers of massive spheroids and active
 galaxies, are considered to play an important role in the formation and the evolution
 of galaxies. It is suggested that SMBHs regulate the star formation activities of
 galaxies through their enormous energy output during active phase, providing a feedback 
 mechanism to reconcile the observed trend of downsizing of the galaxy evolution with hierarchical
 galaxy formation models \citep{juneau05,bundy06,schawinski06}.
  However, the coevolution of SMBHs and their host galaxies 
 stand as an unsolved astrophysical problem, as such a process needs to eventually lead 
 to a rather unexpected tight correlation between host galaxy properties
 and SMBH mass today \citep{ferrarese00,gebhardt00,tremaine02}.

  One of the great challenges toward the understanding of the 
 evolutionary sequence of AGNs and the co-evolution of
 SMBHs and their host galaxies comes from the so-called red or dusty AGNs which
 are believed to occupy more than 50$\%$ of the AGN population \citep{comastri01,tozzi06,polletta08}.
  Red, dusty AGNs are expected, if the initial phase of AGN activity starts in a dust-enshrouded
 environment such as within massive starburst regions of luminous infrared galaxies.
  In such a scenario, AGNs become more visible after sweeping away cold gas and dust that are 
 necessary to sustain the massive star formation in their host galaxies \citep{hopkins06}. 
  On the other hand, dust obscuration can arise in a different way in the unified model of AGN 
 when accretion disks and broad line regions (BLRs) around SMBHs are viewed through dust torus.
  In any case, red, dusty AGNs can shed light on the properties and the evolution of 
 the AGN population in general.

  The phenomenological definition of ``red, dusty AGNs'' is rather broad. It can include 
 AGNs selected in a variety of ways such as those selected by very red colors in optical
 through NIR and the radio detection 
 (e.g., $R-K > 5$ mag and $J-K > 1.3$ mag of the sample in Urrutia et al. 2009; 
 also see Cutri et al. 2001), 
 red MIR colors (Lacy et al. 2004; Lee et al. 2008), and hard X-ray detections (e.g.,
 Polleta et al. 2007).
  These different kinds of AGNs have a common characteristic where their SEDs
 are red, due to the obscuration of their light by the foreground
 gas and dust. Hence, they are considered to be the intermediate population
 between the dust-enshrouded star forming galaxies and the unobscured AGNs.
  The dust-obscuration does not necessarily exclude Type-1 AGNs (AGNs with broad 
 emission lines; e.g., Alonso-Herrero et al. 2006), as red, dusty AGNs with broad emission
 lines are found quite often. Of 23 X-ray QSOs with IR power-law SED and spectroscopic  
 redshift identification, 14 are classified as broad-line AGNs (Szokoly et al. 2004;
 Alonso-Herrero et al. 2006).  More than 50\% of the red AGNs with the 
 radio and red optical/NIR color selection are found to be type-1 AGN which 
 can have the reddening parameter of 2 or more (Uruttia et al.
 2009; Glikman et al. 2007).
  The existence of broad line AGNs among red, dusty AGNs opens a possibility of 
 studying the properties of the dust-obscured quasars in more detail using traditional
 type-1 AGN diagnostics.

  However, even the measurement of the most basic AGN parameters such as $M_{\rm BH}$ and
 the Eddington ratio (the accretion rate) is a difficult task for the red, dusty AGNs.
  In general, $M_{\rm BH}$ are derived from the optical or
  the ultraviolet (UV) part of  AGN spectra for which spectral diagnostics are well established. 
  In the case of dusty AGNs, the dust obscures or significantly reduces the UV or the optical light 
 coming from the region around SMBHs, making the popular AGN optical/UV spectroscopic diagnostics
 useless.    For example, popular $M_{\rm BH}$ estimators are based upon a virial relation
 between two parameters
 - the velocity width of the broad, H$\alpha$ or H$\beta$ lines, and the size of BLR estimated
 from the continuum luminosity at 5100 \AA ${}$ \citep{kaspi00} or the luminosities
 of H$\alpha$ or H$\beta$ \citep{greene05}.
  If the light from a red, dusty AGN is extincted by a color excess of $E(B-V)$ =2 mag 
 \citep{glikman07,urrutia09},
 its H$\alpha$ and H$\beta$ line fluxes would be suppressed by a factor of 100 and 1000 respectively.
 One can try to estimate the amount of the dust extinction from a continuum fitting of the
 optical-UV spectrum or through the Balmer decrement, but such estimates are often
 inconsistent with each other and accompanied with uncertainties of order of $\delta E(B-V) \sim 
 0.5$ mag or more which are related to the dispersion in the intrinsic properties of AGNs.
  The dust obscuration can arise from the Galacitic extinction for AGNs at low galactic
 latitude (e.g., Im et al. 2007; Lee et al. 2008).

  The problem can be substantially alleviated if we can use NIR lines instead of
 optical or UV lines, since NIR Hydrogen lines such as the P$\alpha$ and 
 P$\beta$ are much less affected by the dust extinction than the UV/optical
 light. For the red, dusty AGN with the color excess of $E(B-V)$ = 2 mag,
 the line fluxes of P$\alpha$ and P$\beta$ 
 are suppressed by a factor of only 2.3 and 4.7 respectively. This is a
 significant improvement over the optical lines. Since 
 $M_{\rm BH} \propto L^{0.5}$, where $L$ is a luminosity of the continuum or an emission line, 
 the suppression in the Paschen line luminosities
 introduces uncertainties in $M_{\rm BH}$ estimates only at the level of a factor of 2
 or less, even without correcting for the dust extinction with
 the Paschen decrement.

 In this paper, we will derive the $M_{\rm BH}$ estimators based on the NIR Hydrogen lines
 of Type-1 AGNs and investigate the line ratios of the Paschen lines, keeping in mind 
 future applications of such relations to studies of dusty, red AGNs
 with broad emission lines.

\section{DATA}

\subsection{The Sample}

  In order to construct mass estimators based on the NIR Hydrogen lines, we used two
 samples of low redshift Type-1 AGNs with NIR spectra at 0.8--4.1 $\mu$m. One is
 from Landt et al. (2008; hereafter L08) who studied 23 well-known Type-1
 AGNs in the local universe.
  Among these, we use 16 and 21 AGNs that have line flux and width measurements 
 in P$\alpha$ or P$\beta$ lines, respectively.
  Note that we excluded four objects,
 Mrk 590, NGC 5548, Ark 564, and NGC 7469 in all or some of our analysis even
 though P$\alpha$ and P$\beta$ line flux and  FWHM measurements were available.
 For Ark 564 and NGC 7469, the FWHM values of P$\alpha$ and  P$\beta$ lines listed
 in L08 are too discrepant from each other differing by more than a factor of 1.5.
 This suggests that one of the two measurements is erroneous. Therefore, these
 two objects are excluded from all of the analysis performed below.
  For Mrk 590 and NGC 5548, the H$\alpha$ and the H$\beta$ 
 widths listed in L08 differ by more than a factor of 1.5, but the Paschen
 line widths are consistent with each other. These two objects are excluded  
 in the analysis of the correlation between the Paschen and the Balmer lines, but
 their reverberation-mapping derived $M_{\rm BH}$ are retained for the derivation 
 of the black hole estimators (of the method 2 and the method 3 in Section 3.2). 

  Another sample comes from Glikman et al. (2006; hereafter G06). The G06 sample is made
 of 26 Type-1 AGNs that were selected from a cross-match between the SDSS-DR1
 quasar catalog \citep{schneider03} and the 2 micron All Sky Survey (2MASS;
 Skrutskie et al. 2006), with the criteria of K $<$ 14.5 mag, redshift z $<$ 0.5 and
 the absolute magnitude in $i$-band, $M_{i}$ $<$ -23 mag. Here, the absolute magnitude limit
 is imposed to minimize
 contamination of the quasar light due to the host galaxy. The G06 sample provides
 11 AGNs with P$\alpha$ lines and 14 AGNs with P$\beta$ lines showing up in their
 NIR spectra.

  In all, we use 27 and 35 Type-1 AGNs
 for the P$\alpha$ and the P$\beta$ line analysis, respectively.
 Table \ref{tbl1}
 summarizes properties of these AGNs.

\subsection{Black Hole Mass and Analysis of Optical Spectra}

  For $M_{\rm BH}$ of our sample AGNs, we use the following values.

  First, we use $M_{\rm BH}$ derived with the reverberation mapping method from literatures if
 available.
  This applies to 10 AGNs with P$\alpha$ and 11 AGNs P$\beta$ data in L08
 for which $M_{\rm BH}$ are taken from Vestergaard \& Peterson (2006).

  Second, if $M_{\rm BH}$ from the reverberation mapping method are not available,
 we derived $M_{\rm BH}$ from the optical spectral information using 
 single epoch $M_{\rm BH}$ estimators.
  For the L08 sample without the reverberation mapping derived $M_{\rm BH}$,
 we used the H$\beta$ luminosity and FWHM listed in the Table 5 of L08
 to derive their $M_{\rm BH}$. For this,
 we adopted a $M_{BH}$ estimator in Greene \& Ho (2005), after
 adjusting their relation to have the virial factor of 5.5 \citep{onken04, woo10} by
 multiplying their relation by a factor of 1.8. We also made a small correction
 to the L08 FWHM and luminosity values, since the way how Greene \& Ho (2005) derived
 FWHMs and luminosities is different from what L08 did - Greene \& Ho (2005) adopted 
 a multiple Gaussian component fitting to the broad line while L08 adopted a single
 component fitting of the broad line after removing narrow line components. 
  We find that the correction factor to be 
 FWHM$_{multi}$/FWHM$_{single}$=0.9 and
 $L_{multi}$/$L_{single}$=1.08 on average 
 through fitting of the SDSS spectra of the G06 sample
 with the two methods (See below).

  For the G06 sample, none of the objects have $M_{\rm BH}$ derived from the
 reverberation mapping method. Some of the G06 sample have their H$\beta$ line
 luminosities and FWHMs listed in Shen et al. (2008), but after checking 
 FWHM values in Shen et al. (2008) and the optical spectra, 
 we find cases where FWHMs appear to be overestimated.
  Therefore, we used our own spectral fitting routine (Kim et al. 2006) to obtain
 the H$\beta$ FWHMs and the H$\beta$ luminosities from the SDSS spectra.
  The fitting model includes components for a power law continuum, the host galaxy
 light, broad Fe II multiplet emissions, a Balmer continuum, and multiple Gaussian
 profiles for broad and  narrow emission lines. 
  The FWHM and the line luminosities are derived from the best-fit model profiles 
 of the broad line component which can be consisted of a sum of multiple Gaussian 
 components. 

  From the derived luminosity and FWHM values, we estimated $M_{\rm BH}$ by using
 an updated version of the H$\beta$-based virial mass estimator (Greene \& Ho 2005).
  We updated Equation (7) in Greene \& Ho (2005), by adopting a newer 
 $R_{\mathrm{BLR}}$-$L_{5100}$ relation, i.e., Equation (2) of \citep{bentz09}.
  The updated relation is given below.

 \begin{equation}
\frac{M}{M_{\odot}}=10^{6.88\pm0.57}(\frac{L_{\mathrm{H\beta}}}{10^{42}~\mathrm{erg~  s^{-1}}})^{0.46\pm0.05}(\frac{\mathrm{FWHM_{H\beta}}}{10^{3}~\mathrm{km~s^{-1}}})^{2}.
 \end{equation}

  Again, a proper adjustment was 
 made for the virial factor in order to make the relation 
 consistent with the virial factor in Vestergaard \& Peterson (2006).
  The derived BH masses are listed in Table 1.

\subsection{Analysis of NIR Spectra and Line Information}

   The G06 data were taken mostly with the SpeX instrument on the IRTF with the spectral resolution
 of $\frac{\lambda}{\Delta \lambda} \simeq 1200$ at 0.8 - 2.5 $\mu$m, and 
  $\frac{\lambda}{\Delta \lambda} \simeq 1500$ at 1.9 - 4.1 $\mu$m.
  G06 provide with the flux-calibrated, reduced NIR spectra,
  and we used the following procedure to measure the Paschen line fluxes and FWHMs
  of the G06 sample.

  The G06 spectra were converted to the rest-frame, and then the continuum around the Paschen
 lines were determined by fitting a linear function to the continuum at regions near
 the Paschen lines. We varied the wavelength centers and widths of the continuum
 fitting regions,  
  but typical values are chosen to be $\sim$2 FWHM of the broad component of
 H$\beta$ line away from the Paschen line center 
 and the width of 0.02 $\mu$m for the continuum fitting region centers
 and widths respectively.  For the P$\alpha$ line, this corresponds to 
 the wavelength regions centered somewhere at 1.82 to 1.84 $\mu$m and 1.90 to 1.92 $\mu$m
 with the width of 0.02 $\mu$m, while for the P$\beta$ line
 the wavelength centers are somewhere at 1.23 to 1.25 $\mu$m and 1.31 to 1.33 $\mu$m.
  When the continuum shows data points affected by imperfect sky subtraction,
 we adjusted the continuum fitting regions. Also,  
  the width of the continuum fitting regions are varied from 0.01 to 0.03 $\mu$m,
 and the center wavelengths are shifted by 0.01-0.02 $\mu$m to provide robust measurements.
  Although  NIR spectra of Type-1 AGNs are known to be dominated by
 a black body radiation from warm dust, we find that the linear fitting of the
 continuum over the limited wavelength range provides a reasonable
 approximation of the continuum around the Paschen lines (see Figure 2; also see L08).

  After the continuum subtraction, we used a single Gaussian function to fit an emission
 line. The fitted parameters are multiplied by correction factors to correct for
 systematic errors in the single component fitting (see next paragraph).
  During the fitting, we set free the central wavelength of each Gaussian component. 
  Therefore, the free parameters of the fit are FWHM, flux, and the
 central wavelength for Gaussian component.
  The fit was performed with the MPFITEXPR procedure
 of the IDL. Figures 2 and 3 show the line-fitting results.
  The fits provide a formal error of 12$\%$ in flux and 9$\%$ in FWHM.
  We also estimated the error arising from uncertainties related to how the continuum is determined.
 This was done by varying the regions for the continuum fitting as described
 earlier. We find that the average uncertainty
 arising from the continuum determination is 4$\%$ (flux), 3$\%$ (FWHM), although
 we find cases where the FWHM values change by as much as 25$\%$. 
 Such cases occur when there is an extended feature at the wing of the emission line
 which may or may not be a true feature. 
  The total errors in flux and FWHM measurement are taken as
 a square root of the quadratic sum of the formal fitting error
 and the error from the continuum determination, which are found to be typically
 12\% and 9\%.

  Fitting a multiple component line with a single Gaussian function can produce 
 systematic errors in the fitted parameters, and we tried to correct for such a bias
 by applying correction factors that are derived from Balmer lines. This is done by comparing the broad
 line widths and the line  fluxes from multiple component fits of Balmer lines versus the fitted parameters from 
 a single component fit of the same Balmer lines.
  By doing so, we correct potential systematic arising from not removing narrow line components 
 in the fitting process too. The correction factors are computed for each object and they range 
 over 0.74 - 1.01 with the mean of 0.91 for the FWHM ratios (FWHM$_{multi}$/FWHM$_{single}$) and 
 0.98 - 1.17 for the line flux ratio with the mean of 1.06 ($L_{multi}$/$L_{single}$). 
  This procedure makes an assumption that the line profile shapes of the Paschen lines follow  
 those of the Balmer lines.
  To confirm this assumption, we compared the P$\beta$ line profiles with the model-fitted line
 profiles of H$\alpha$. Only the P$\beta$ lines with the S/N $>$ 25 are shown in Figure 4.
  The P$\beta$ lines are plotted as histograms with error bars, while the H$\alpha$ lines are 
 indicated with thick solid lines. The thin dotted lines and the thick dashed lines 
 represent the broad and the narrow components of the H$\alpha$ lines.  
 The P$\beta$ profiles are normalized to have the maximum value of 1 while the H$\alpha$ are normalized
 to have the same total flux as the normalized P$\beta$ profiles.
 Figure 4 demonstrates that the P$\beta$ profile shapes are consistent with those of the H$\alpha$ profiles,
 except for the width and the overall normalization, supporting our assumption 
 that the Paschen line profiles are similar to those of the Balmer lines.
  We also tried fitting the Paschen line profiles with double Gaussian profiles for objects
  that seem to have narrow line components
 (SDSSJ010226.3$-$003904.6, SDSSJ015530.0$-$085704.0, SDSSJ015950.2$+$002340.8, SDSSJ031209.2$-$081013.8,
  SDSSJ015910.0$+$010514.5, SDSSJ235156.1$-$010913.3, SDSSJ000943.1$-$090839.2 and SDSSJ011110.0$-$101631.8).
  For H$\beta$, we find that the double Gaussian fitting returns FWHM values greater than those from the 
 single Gaussian fitting by a factor of 1.08 in average. 
  However, the double Gaussian fitting (one for narrow line, and another 
 for broad line) tends to return FWHM values greater than the results from the multiple component fitting. 
  Thus, another correction factor is needed to convert the double Gaussian fitting result to the multiple
 Gaussian fitting result, and the result from this double Gaussian fitting test should be taken
 seriously, We are presenting a result from this test in order to see if we find the same kind of  
 the dependency of fitted parameters on the fitting method which we find in the Balmer lines as a way to
 support our assumption that the Paschen line profiles are similar to the Balmer line profiles.

  In most case, the fitted values from the double component fit agree with those from the single component
 fit (without the additional correction factor), where the FWHM and flux values from the
 double component fittings are in average  7\% larger and 2\% smaller than the single component fitting results
 respectively.  This agrees well with the trend we find from our analysis of the Balmer lines 
 using single, double, and multiple component Gaussian fittings (see above).
  One exception is SDSSJ015950.2$+$002340.8 whose FWHM changed by about 50\%. Distinguishing
 the narrow and the broad components is difficult for this object, therefore  
 the mean value between the parameters from the two different methods
 was adopted as the best-fit value, with the half of the difference as its error.
  We note that the double component fit improved the reduced $\chi^{2}$ values significantly ($> 0.3$)
 for only three objects, where the improvement came from
 fitting of broad extended wings which did not affect the derived fitting parameter 
 values rather than through the change of the FWHM values (except for SDSSJ015950.2$+$002340.8).

  Finally, the measured FWHMs were corrected for the instrumental resolution
 and the fitting methodology (multiple component fit versus single component fit),
 and the fluxes were converted to the luminosity assuming a standard $\Lambda$CDM cosmology
 of $H_{0}$=70 km sec$^{-1}$ Mpc$^{-1}$, $\Omega_{m}$=0.3 and $\Omega_{\Lambda}$=0.7
 (e.g., Im et al. 1997).
  The measured line luminosities, fluxes, and FWHMs are presented in Table \ref{tbl1}.

  L08 present the results of their line analysis, based on the NIR spectra that
 were taken with the Spex spectrograph \citep{rayner03} on the Infrared Telescope
 Facility (IRTF) at an average spectral resolution of 400 $\mathrm{km~s^{-1}}$. 
  For the L08 sample, we use the line fluxes and FWHMs derived by them
 and presented in their Table \ref{tbl2}, after correcting FWHM values for 
 the instrumental resolution. We also applied the correction factor of 
 FWHM$_{multi}$/FWHM$_{single broad}$=0.9 and $L_{multi}$/$L_{single broad}$=1.08 which 
 corrects for the difference in the line-fitting methods (L08 versus Greene \& Ho 2005) 
 to derive the line parameters which are similar to the correction factors 
 we derived for the fitting process of the G06 spectra.
 As for the accuracy of the L08 measurements, L08 quote
 a typical error of 10\% of less. Therefore, we adopt a conservative value of 10\% for
 the measurement error of the fluxes and the FWHMs presented in L08.

  We note that contamination from the host galaxy light to these measurements is negligible.
  The contamination of the line flux due to the host galaxy is possible, but the G06 data were
 taken with a narrow slit to minimize the host galaxy light to less than 8\%. 
  We expect that the same statement holds true for the L08 sample whose data were taken 
 with a narrow slit for AGNs at $z < 0.1$ which are located much closer to us
 than the G06 sample.

\section{RESULTS}

\subsection{Empirical Relation between Balmer and Paschen Lines}

  Before constructing mass estimators based on the Paschen lines, we show here 
 that how well the properties of the Paschen lines correlate with the Balmer lines.
 A tight correlation
 between the two lines would imply that the Paschen lines
 originate from the broad line regions similar to the Balmer lines, thus
 serving as a strong justification for the use of the Paschen lines as a mass estimator.
  Good correlations between the FWHM values of the broad Balmer lines and the broad Paschen lines
 were shown in L08.
  Here, we extend the analysis to the line flux ratios, and add quasars from the G06 sample
 to the L08 sample to strengthen the goodness of the FWHM correlation.
  Furthermore, we also derive equations that relate the properties of the Balmer and
 the Paschen lines.
  Figure 5 shows the correlation between FWHM values of the Balmer and the Paschen broad lines,
 while Figure 6 shows a correlation of line luminosities 
 of the broad lines.
  To derive the correlations between the two quantities,
 we performed a linear bisector fit using the equations below. 

\begin{equation}
\log(\frac{\mathrm{FWHM_{Y}}}{1000~\mathrm{km~s^{-1}}})=\mathrm{A+B}\log(\frac{\mathrm{FWHM_{X}}}{1000~\mathrm{km~s^{-1}}})
\end{equation}

\begin{equation}
\log(\frac{L_{\mathrm{Y}}}{10^{42}~\mathrm{erg~s^{-1}}})=\mathrm{C+D}\log(\frac{L_{\mathrm{X}}}{10^{42}~\mathrm{erg~s^{-1}}})
\end{equation}

 Here,  X and Y are line identifiers, and  A and B are the correlation coefficients
 in the fit for FWHM, and
 C and D are the coefficients for the line luminosity ratio fit.
 The results of the fitting are summarized in Table 2. The table also lists the 
 rms scatter of the data points with respect to the best-fit lines.

  These results show that the Paschen line luminosities and
 FWHMs correlate well with those of the Balmer lines.
  The rms scatters in the line luminosity correlation 
 are $\sim$0.12-0.14 dex against H$\alpha$, and $\sim$0.16-0.19 dex against H$\beta$.
 For the FWHM values, the rms scatters are 0.045-0.06 dex against H$\alpha$, and 0.05-0.06 dex
 against H$\beta$. The slightly larger scatters 
 and a notable offset in FWHM values of H$\beta$ against Paschen lines suggest
 the complexities in AGN spectra around H$\beta$ line noted in L08,
 an excess, broad component in the red part of the H$\beta$ line caused by
 an unclear origin (e.g., Meyers \& Peterson 1985; V{\'e}ron et 
al. 2002).
  We also point out that the line widths of Balmer lines are systematically larger 
 than those of the Paschen lines, and that the trend is stronger as the wavelength 
 decreases. This suggests that the Paschen broad lines and the Balmer broad lines
 originate from a similar BLR, but with Balmer lines originating from 
 the inner region of the BLR than Paschen lines.

  Similarly, we also examine correlations between ($L_{5100\mathrm{\AA{}}}$) and
 $L_{\mathrm{P\alpha,\beta}}$. 
  $R_{\mathrm{BLR}}$ values are derived from $L(5100)$ using Equation (4) of \citep{greene05}.
  Figure 7 shows the correlation between $R_{BLR}$ and Paschen line luminosities, and
 Equations (4) and (5) are the best-fit results.

 \begin{equation}
R_{\mathrm{BLR}}=(50.5\pm1.0){(\frac{L_\mathrm{{P\alpha}}}{10^{42}~\mathrm{erg~s^{-1}}})}^{(0.40\pm0.01)}~\mathrm{lt-days},
 \end{equation}

 \begin{equation}
R_{\mathrm{BLR}}=(50.6\pm1.0){(\frac{L_\mathrm{{P\beta}}}{10^{42}~\mathrm{erg~s^{-1}}})}^{(0.48\pm0.01)}~\mathrm{lt-days}.
 \end{equation}

  Figure 7 and the above result indicate that the luminosity of the Paschen lines
 is closely connected to the radius of BLR.

   These tight correlations support the idea that
 the Paschen line emissions originate from the BLR region similar to the area
 where the broad lines of the Balmer lines arise. Thus, it should be 
 possible to derive $M_{\rm BH}$ from P$\alpha$ and P$\beta$ just like
  $M_{\rm BH}$ can be estimated from H$\beta$ and H$\alpha$.
  Now, we move on to the derivation 
 of $M_{\rm BH}$ estimators based on Paschen lines.

\subsection{Mass Estimators}

  In this section, we derive new $M_{\rm BH}$ estimators based on P$\alpha$ and
 P$\beta$ lines over the mass range of $10^{6.4}~\sim~10^{9.5}~M_{\odot}$.
  These $M_{\rm BH}$ estimators are analogous to the $M_{\rm BH}$ estimators based on
 H$\alpha$ and H$\beta$ lines (Greene \& Ho 2005),
 with  $L_\mathrm{{P\alpha}}$ and $L_\mathrm{{P\beta}}$ serving as a measure of
 the BLR size, and FWHMs of the Paschen lines as the velocity term in the 
 virial mass estimator. Mathematically, we need find three unknown parameters,
 $a$, $b$, and $c$ in the following equation.

\begin{equation}
 \log(M)= \mathrm{a + b}\,\log(L)+\mathrm{c}\,\log(\mathrm{FWHM})
\end{equation}

  The expected values are $\mathrm{c}=2$ from the virial theorem, and $\mathrm{b}=0.5$
 if $L\sim$ $R_\mathrm{{BLR}}^2$ (Dibai 1977).
  Since there are several ways to derive $M_{\rm BH}$ estimators and the results
 may not be identical, 
  we derive $M_{\rm BH}$ estimators in three different ways as described below.

 First, we derive $M_{\rm BH}$ estimators by simply replacing the line luminosities and FWHMs
 of the Hydrogen Balmer line mass estimators (A1) of \citep{greene07}
 with those of the Paschen lines using the relation presented in Section 3.1.
 We used the H$\alpha$-based $M_{\rm BH}$ estimator
 since the comparison of H$\alpha$ and the Paschen line properties
 show smaller scatter values than H$\beta$. We replaced the H$\alpha$ luminosity and
 FWHM in the base estimator with those of P$\alpha$ or P$\beta$ lines. 
  A factor of 1.8 is multiplied to the H$\alpha$-based $M_{\rm BH}$ estimator of 
 Greene \& Ho (2005) so that the new relation is normalized to the virial factor of $f=5.5$.
  The $M_{\rm BH}$ values for the left-hand side of Equation (6) are those listed in Table \ref{tbl1}. 
  The Paschen line based $M_{\rm BH}$ estimators derived this way are given below.

\begin{equation}
\frac{M}{M_{\odot}}=10^{7.29\pm0.10}(\frac{L_{\mathrm{P\alpha}}}{10^{42}~\mathrm{erg~ s^{-1}}})^{0.43\pm0.03}(\frac{\mathrm{FWHM_{P\alpha}}}{10^{3}~\mathrm{km~s^{-1}}})^{1.92\pm0.18}
\end{equation}
\begin{equation}
\frac{M}{M_{\odot}}=10^{7.33\pm0.10}(\frac{L_{\mathrm{P\beta}}}{10^{42}~\mathrm{erg~ s^{-1}}})^{0.45\pm0.03}(\frac{\mathrm{FWHM_{P\beta}}}{10^{3}~\mathrm{km~s^{-1}}})^{1.69\pm0.16}
\end{equation}

 Figure 8 shows $M_{\rm BH}$ from this estimator versus the input $M_{\rm BH}$ values.
 The estimators provide a reasonable fit with rms scatter of  0.22, 0.23
 dex for both $M_{\rm BH}$ (P$\alpha$) and $M_{\rm BH}$ (P$\beta$).

 The second and the third methods utilize a direct fit of the $M_{\rm BH}$ values.
 In the second method, we fix the exponent of the velocity term c to 2 as expected
 in the virial relation, while setting the overall normalization, a, and the exponent
 of the luminosity term, b, as free parameters.
  For this, we performed a linear regression fit in logarithmic scale as in 
 Equation (6), using the REGRESS procedure in IDL.

  For the second method, we obtain the following relation:

\begin{equation}
\frac{M}{M_{\odot}}=10^{7.16\pm0.04}(\frac{L_{\mathrm{P\alpha}}}{10^{42}~\mathrm{erg~ s^{-1}}})^{0.49\pm0.06}(\frac{\mathrm{FWHM_{P\alpha}}}{10^{3}~\mathrm{km~s^{-1}}})^{2.}
\end{equation}
\begin{equation}
\frac{M}{M_{\odot}}=10^{7.13\pm0.02}(\frac{L_{\mathrm{P\beta}}}{10^{42}~\mathrm{erg~ s^{-1}}})^{0.48\pm0.03}(\frac{\mathrm{FWHM_{P\beta}}}{10^{3}~\mathrm{km~s^{-1}}})^{2.}
\end{equation}

 In Figure 9, we compare the input $M_{\rm BH}$ versus $M_{\rm BH}$ from the Paschen estimators of
 the 2nd method. We find that the estimators in Equation (9) and (10) 
 reproduce $M_{\rm BH}$ with rms scatters of about 0.20-0.24 dex, not much from the method 1.

 In the third method, we treat all the coefficients in Equation (6) as free parameters.
 To derive the coefficients in the relation, we use a multiple variable linear
 regression fit with  the REGRESS fitting procedure in IDL.

  For the third method, we obtain the following relation:

\begin{equation}
\frac{M}{M_{\odot}}=10^{7.31}(\frac{L_{\mathrm{P\alpha}}}{10^{42}~\mathrm{erg~ s^{-1}}})^{0.48\pm0.03}(\frac{\mathrm{FWHM_{P\alpha}}}{10^{3}~\mathrm{km~s^{-1}}})^{1.68\pm0.12}
\end{equation}
\begin{equation}
\frac{M}{M_{\odot}}=10^{7.40}(\frac{L_{\mathrm{P\beta}}}{10^{42}~\mathrm{erg~ s^{-1}}})^{0.46\pm0.02}(\frac{\mathrm{FWHM_{P\beta}}}{10^{3}~\mathrm{km~s^{-1}}})^{1.41\pm0.09}
\end{equation}

  Figure 10 shows the comparison of the input $M_{\rm BH}$ versus $M_{\rm BH}$ from the estimators in Equations (11) and (12).
  The rms scatter in the derived $M_{\rm BH}$ is reduced to 0.19, 0.21 dex while the exponents
 of the velocity term are found to be about 1.4-1.7 rather than the value of 2 expected in a virial relation.
  Compared to the 2nd method, the third method improves the fitting accuracy by a small amount
 in terms of reducing the rms scatter by 0.03 dex.

  Also, note that the scatter in the $M_{\rm BH}$ estimator is intrinsic (e.g., the uncertainty 
 in the input $M_{\rm BH}$ values) rather than dominated by the measurement errors 
 in the FWHM and line luminosity measurements. The estimated 1-$\sigma$ errors in FWHM and
 line luminosity measurements are 0.04 dex and 0.05 dex respectively, and such errors can 
 produce a scatter in the $M_{\rm BH}$ estimators at the level of 0.09 dex only.

\section{DISCUSSION}

\subsection{Consistency between three $M_{\rm BH}$ estimators}

  We derived the $M_{\rm BH}$ estimators in three different ways. 
  Ideally, each estimator should give a $M_{\rm BH}$ value consistent with the others 
 within the intrinsic rms scatter for a given set of realistic FWHM and
 line luminosity.
  To see if there is any noticeable deviation among
 the derived BH mass values from the three different estimators,
  we compare 
 $M_{\rm BH}$ values derived from the three estimators. 
  For the input sets of FWHM and line luminosities, we 
 convert the available H$\beta$ FWHM and  $L(5100)$ of SDSS quasars from Shen et al. (2008)
 to the Paschen line quantities,
 which gives realistic sets of FWHM and line luminosity values in a large range
 of FWHM and luminosity parameter space occupied by quasars.
  To convert FWHM$_{\mathrm{H\beta}}$ to FWHM$_{\mathrm{P\alpha}}$ or FWHM$_{\mathrm{P\beta}}$ and
 $L(5100)$ to $L_{\mathrm{P\alpha}}$ and $L_{\mathrm{P\beta}}$, we use the 
 the empirical relations presented in Section 3.1 of this paper. The derived 
 Paschen line FWHM and luminosities are fed into the new $M_{\rm BH}$ estimators.

  Figure 12 shows the comparison of $M_{\rm BH}$ derived from three different estimators.
 The left panel of the figure shows the comparison between $M_{\rm BH}$ values from the method 1 against 
 those from the method 2.
 There is a small systematic offset of 0.1, 0.15 dex in the derived
 $M_{\rm BH}$ values at $M_{\rm BH} \sim 10^{10.0}~M_{\odot}$, $M_{\rm BH} \sim 10^{7.0}~M_{\odot}$
 with the overall rms scatter of 0.06 dex.
 Considering that the rms scatter in these two estimators is about 0.2 dex,
 these two estimators produce $M_{\rm BH}$ values well within the rms scatter in their relation.
  The right panel of the Figure 12 compares $M_{\rm BH}$ from the method 2 versus the method 3.
 There is a significant systematic offset of 0.2 dex at the high mass end of $10^{10}~M_{\odot}$
 and the low mass end of $10^{7}~M_{\odot}$ -- both are comparable the rms scatter of the $M_{\rm BH}$ estimators.
  The systematic offset results mainly from the difference in the exponent of FWHM in
 the $M_{\rm BH}$ estimators.
  The derived $M_{\rm BH}$ is proportional to $\mathrm{FWHM^{2}}$ in the method 2, 
 while $M_{\rm BH}$ from the method 3 is proportional to $\mathrm{FWHM^{1.5}}$.
 Therefore, the method 2 is bound to produce $M_{\rm BH}$ values larger than the method 3
 at large $M_{\rm BH}$ while the opposite trend appears at low $M_{\rm BH}$.

  It is difficult to determine which method gives the best estimate of $M_{\rm BH}$.
  The method 3 gives the smallest scatter, but the FWHM exponent of 1.5 is less physical
 than the index of 2 in the method 2.
  The estimator from the method 3 may be proved to be
 the best choice, if a good explanation can be provided why the exponent of FWHM is 1.5, not 2. 
  The $M_{\rm BH}$ estimator from the method 3 has 
 a resemblance to the fundamental plane relation of early-type galaxies (Dressler et al. 1987; Djorgovski 
 1987) where it still remains controversial why the relation is slightly tilted off from the
 virial relation (e.g., see Jun \& Im 2008), so a similar physical mechanism that produces the tilt in 
 the fundamental plane might be in work for the $M_{\rm BH}$ estimator.
  For now, we prefer the method 2 the most simply because the formula can be
 justifiable easily on a physical basis. 

\subsection{Line Ratios}

 The line ratios of H$\beta$ through P$\alpha$ lines can give us clues on the physical conditions
 of the BLR region.
 The line ratios can also serve as a basis for determining the extinctions at different wavelengths
 in BLR of dusty AGN. Therefore, we present the line ratios of the Hydrogen lines of our sample.   

 Figure 11 shows the line ratios of H$\beta$, H$\alpha$, P$\beta$ to P$\alpha$
 of our sample. Overplotted are theoretical
 line ratios based on the computation from the CLOUDY code \citep{ferland98}.
  Using the CLOUDY code, we calculate the expected line ratios by varying three parameters,
 the shape of the ionizing continuum ($\alpha$=-1.0, -1.5), the ionization parameter
 ($U= 10^{0.5}$, $10^{-0.5}$, $10^{-1.5}$) and the hydrogen density ($n= 10^{9}$, $10^{10}$,
 $10^{11}$~$\mathrm{cm^{-3}}$).
  We find that the observed line ratios are reproduced most successfully with a set of 
 parameters, $\alpha= -1.0$, $U= 10^{-1.5}$ and $n= 10^9~\mathrm{cm^{-3}}$. Some line ratios that
 are below the theoretical expectation can be explained naturally if there is a small
 amount of extinction with color excess of $E(B-V) < 0.4$ mag.
  The mean line ratios are found to be $\mathrm{\frac{H\beta}{P\alpha}} = 2.70$,  $\mathrm{\frac{H\alpha}{P\alpha}} = 9.00$,  
 and $\mathrm{\frac{P\beta}{P\alpha}} = 0.91$. These values can be used as a basis for determining extinctions
 in the BLR of dusty AGN.  
 The perfect fit of the line ratios as a function of wavelengths is probably not possible with
 a simple model, considering that the broad emission lines at different wavelengths are not 
 likely to originate from exactly the same BLR regions as we discussed in Section 3.1.
  Nevertheless, the observed line ratios are consistent with the expected ratios from a single set
 of ionization parameters.

\section{CONCLUSION}

  We derived new $M_{\rm BH}$ estimators, using Hydrogen P$\alpha$ and P$\beta$ lines of Type-1 AGNs.
  The derived estimator allows the determination of $M_{\rm BH}$ at the accuracy of $\sim 0.2$ dex,
 and they will be useful for estimating $M_{\rm BH}$ of red, dusty AGNs.
  Our analysis of the Paschen lines with respect to H$\alpha$ and H$\beta$ lines
 shows that the luminosities and FWHMs of the broad components of the Paschen lines correlate well with 
 those of the Balmer lines. The Hydrogen line ratios from H$\beta$ through P$\alpha$ are consistent
 with a Case B recombination with
 the parameters of $\alpha= -1.0$, $U= 10^{-1.5}$ and $n= 10^9~\mathrm{cm^{-3}}$.
 The mean line ratios are $\mathrm{\frac{H\beta}{P\alpha}} = 2.70$,
 $\mathrm{\frac{H\alpha}{P\alpha}} = 9.00$, and $\mathrm{\frac{P\beta}{P\alpha}} = 0.91$
 which can be used to estimate for the amount of 
 dust extinction present in red AGNs in future.
 Future applications of these results on red, dusty
 AGNs will enable us to better understand the nature of red, dusty AGNs whose properties are
 still hidden behind a wall of dusty gas.

\acknowledgments

 We thank the referee for useful comments which improved the manuscript.
 This work was supported by the grant No. 2009-063616 of the Creative Initiative
 Program, funded by the Korea government (MEST).

\clearpage
\begin{turnpage}
\begin{deluxetable*}{cccccccccccc}
\tabletypesize{\scriptsize}
\tablewidth{0pt}
\tablenum{1}
\tablecaption{Paschen Parameters And BH Masses\label{tbl1}}
\tablehead{
\colhead{AGNs Name}& \colhead{Redshift}& \colhead{BHmass}& 
\multicolumn{4}{c}{P$\beta$}& \colhead{}& \multicolumn{4}{c}{P$\alpha$}\\
\cline{4-7} \cline{9-12} \\
\colhead{}& \colhead{z}& \colhead{log($M/M_{\odot}$)}& 
\colhead{FWHM}& \colhead{$\Delta$FWHM}& \colhead{Flux}&
\colhead{$\Delta$flux}& \colhead{}& 
\colhead{FWHM}& \colhead{$\Delta$FWHM}& \colhead{Flux}&
\colhead{$\Delta$flux} \\
\colhead{}& \colhead{}& \colhead{}& 
\colhead{[$\mathrm{km~s^{-1}}$]}& \colhead{[$\mathrm{km~s^{-1}}$]}& \colhead{[$\mathrm{erg~s^{-1}~cm^{-2}}$]}&
\colhead{[$\mathrm{erg~s^{-1}~cm^{-2}}$]}& \colhead{}& 
\colhead{[$\mathrm{km~s^{-1}}$]}& \colhead{[$\mathrm{km~s^{-1}}$]}& \colhead{[$\mathrm{erg~s^{-1}~cm^{-2}}$]}&
\colhead{[$\mathrm{erg~s^{-1}~cm^{-2}}$]}} 
\startdata
3C273\tablenotemark{a}	& 0.158 & 8.94\tablenotemark{c}& 2608	& --& 6.95E-13	& --& 
	& 2635 	& --& 8.33E-13 & --\\
Mrk876\tablenotemark{a}	& 0.129 & 8.44\tablenotemark{c}& 5433	& --& 7.81E-14	& --&
	& 4965 	& --& 1.31E-13 & --\\
PG0844$+$349\tablenotemark{a}	& 0.064 & 7.96\tablenotemark{c}& 2144	& --& 1.06E-13	& --& 
	& 1967 	& --& 1.21E-13 & --\\
Mrk110\tablenotemark{a}	& 0.035 & 7.40\tablenotemark{c}& 1689	& --& 1.50E-13	& --&
	& 1604 	& --& 1.99E-13 & --\\
Mrk509\tablenotemark{a}	& 0.034 & 8.15\tablenotemark{c}& 2755	& --& 5.20E-13	& --& 
	& 2747 	& --& 6.60E-13 & --\\
Ark120\tablenotemark{a}	& 0.033 & 8.17\tablenotemark{c}& 4610	& --& 5.89E-13	& --& 
	& 4584 	& --& 6.96E-13 & --\\
Mrk817\tablenotemark{a}	& 0.031 & 7.69\tablenotemark{c}& 4988	& --& 1.39E-13	& --& 
	& 4202 	& --& 1.39E-13 & --\\
Mrk335\tablenotemark{a}	& 0.026 & 7.15\tablenotemark{c}& 1633	& --& 1.67E-13	& --& 
	& 1431 	& --& 1.40E-13 & --\\
Mrk590\tablenotemark{a}	& 0.026 & 7.67\tablenotemark{c}& 3565	& --& 1.46E-14	& --&
	& 4259 	& --& 2.82E-14 & --\\
NGC5548\tablenotemark{a}	& 0.017 & 7.82\tablenotemark{c}& 5891	& --& 1.77E-13	& --&
	& 4102 	& --& 1.64E-13 & --\\
NGC4151\tablenotemark{a}	& 0.003 & 7.12\tablenotemark{c}& 4204	& --& 1.21E-12	& --& 
	& --	& --& --	 & --\\
H1821$+$643\tablenotemark{a}	& 0.297 & 9.44\tablenotemark{d}& 4717	& --& 1.92E-13	& --& 
	& -- 	& --& --	 & --\\
PDS456\tablenotemark{a}	& 0.184 & 8.70\tablenotemark{d}& 1828	& --& 3.79E-13	& --&
	& 1785 	& --& 4.70E-13  & --\\
Mrk290\tablenotemark{a}	& 0.030 & 8.09\tablenotemark{d}& 3818	& --& 1.55E-13	& --& 
	& 3180 	& --& 1.64E-13 & --\\
H2106$-$099\tablenotemark{a}	& 0.027 & 7.72\tablenotemark{d}& 2155	& --& 1.70E-13	& --& 
	& 1550 	& --& 2.35E-13 & --\\
Mrk79\tablenotemark{a}	& 0.022 & 7.94\tablenotemark{d}& 3163	& --& 1.86E-13	& --& 
	& 3051	& --& 2.58E-13 & --\\
NGC4593\tablenotemark{a}	& 0.009 & 7.83\tablenotemark{d}& 3407	& --& 2.87E-13	& --& 
	& -- 	& --& --	 & --\\
NGC3227\tablenotemark{a}	& 0.004 & 7.42\tablenotemark{d}& 2680	& --& 3.48E-13	& --& 
	& -- 	& --& --	 & --\\
HE 1228$+$013\tablenotemark{a}	& 0.117 & 7.87\tablenotemark{d}& 1731	& --& 6.62E-14	& --& 
	& 1725 	& --& 7.46E-14 & --\\
H1934$-$063\tablenotemark{a}	& 0.011 & 6.80\tablenotemark{d}& 1241	& --& 1.37E-13	& --& 
	& 1213 	& --& 1.77E-13 & --\\
IRAS1750$+$508\tablenotemark{a}	& 0.300 & 8.33\tablenotemark{d}& 1758	& --& 4.50E-14	& --&
	& -- 	& --& --	& --\\
SDSSJ000943.1$-$090839.2\tablenotemark{b} 	& 0.210	& 8.44\tablenotemark{e}& 4817	& 818& 8.41E-15	&
 1.84E-15&	& 3841	& 724& 8.31E-15 & 2.10E-15\\
SDSSJ005812.8$+$160201.3\tablenotemark{b} 	& 0.211 & 8.29\tablenotemark{e}& 3158	& 409& 1.73E-14	&
 1.96E-15&	& 2775	& 357& 2.23E-14 & 2.37E-15\\
SDSSJ010226.3$-$003904.6\tablenotemark{b} 	& 0.295	& 7.67\tablenotemark{e}& 1565	& 184& 2.42E-14	&
 1.87E-15&	& 1234	& 142& 2.86E-14 & 2.04E-15\\
SDSSJ011110.0$-$101631.8\tablenotemark{b} 	& 0.179	& 8.20\tablenotemark{e}& 3695	& 496& 1.48E-14	&
 1.77E-15&	& 2617	& 380& 1.41E-14 & 1.99E-15\\
SDSSJ015530.0$-$085704.0\tablenotemark{b} 	& 0.165	& 8.85\tablenotemark{e}& 7355	& 1032& 1.53E-14&
 2.38E-15&	& 5074	& 737& 1.36E-14 & 2.24E-15\\
SDSSJ015910.0$+$010514.5\tablenotemark{b} 	& 0.217	& 7.98\tablenotemark{e}& 2657	& 404& 1.01E-14	&
 1.48E-15&	& 2574	& 455& 1.02E-14 & 1.89E-15\\
SDSSJ015950.2$+$002340.8\tablenotemark{b} 	& 0.163	& 8.09\tablenotemark{e}& 2657	& 391& 2.65E-14	&
 2.53E-15&	& 2716	& 627& 3.78E-14 & 3.99E-15\\
SDSSJ021707.8$-$084743.4\tablenotemark{b} 	& 0.292	& 7.84\tablenotemark{e}& 1970	& 552& 3.49E-15	&
 1.14E-15&	& --	& -- & -- & --	 \\
SDSSJ024250.8$-$075914.2\tablenotemark{b} 	& 0.378	& 8.00\tablenotemark{e}& 1329	& 194& 9.28E-15	&
 1.15E-15&	& --	& -- & -- & --	 \\
SDSSJ031209.2$-$081013.8\tablenotemark{b} 	& 0.265	& 8.29\tablenotemark{e}& 3540	& 620& 1.12E-14	&
 2.01E-15&	& --	& -- & -- & --	 \\
SDSSJ032213.8$+$005513.4\tablenotemark{b} 	& 0.185	& 8.03\tablenotemark{e}& 2014	& 234& 2.58E-14	&
 1.89E-15&	& 1894	& 220& 3.40E-14 & 2.47E-15\\
SDSSJ150610.5$+$021649.9\tablenotemark{b} 	& 0.135	& 8.11\tablenotemark{e}& --	& -- & --	&
 --&	& 2632	& 330& 2.26E-14 & 2.54E-15\\
SDSSJ172711.8$+$632241.8\tablenotemark{b} 	& 0.218	& 8.61\tablenotemark{e}& --	& -- & --	&
 --&	& 5087	& 1192& 1.13E-14 & 2.86E-15\\
SDSSJ211843.2$-$063618.0\tablenotemark{b} 	& 0.328	& 8.42\tablenotemark{e}& 2485	& 430& 6.58E-15	&
 1.29E-15&	& --	& -- & -- & --	 \\
SDSSJ234932.7$-$003645.8\tablenotemark{b} 	& 0.279	& 8.19\tablenotemark{e}& 2578	& 360& 1.02E-14	&
 1.47E-15&	& --	& -- & -- & --	 \\
SDSSJ235156.1$-$010913.3\tablenotemark{b} 	& 0.174	& 8.89\tablenotemark{e}& 6137	& 705& 4.42E-14	&
 3.53E-15&	& 4946	& 553& 5.14E-14 & 3.65E-15\\

\enddata
\tablenotetext{a}{FWHM and Flux of these objects come from L08.}
\tablenotetext{b}{FWHM and Flux of these objects measured from spectra presented in G06.}
\tablenotetext{c}{The $M_{\rm BH}$ values are determined from reverberation mapping techniques (Vestergaard $\&$ Peterson 2006).}
\tablenotetext{d}{The $M_{\rm BH}$ values are based on H$\beta$ widths and luminosities in L08.}
\tablenotetext{e}{The $M_{\rm BH}$ values are based on H$\beta$ widths and L(5100) derived from our fit to SDSS
 spectra.}
\end{deluxetable*}
\end{turnpage}
\clearpage

\begin{deluxetable*}{cccccccccc}
\tabletypesize{\scriptsize}
\tablewidth{0pt}
\tablenum{2}
\tablecaption{Coefficients of the Empirical Correlation between Balmer and Paschen Lines\label{tbl2}}
\tablehead{
\colhead{}& \colhead{}& \colhead{}& \multicolumn{3}{c}{FWHM}&
\colhead{}& \multicolumn{3}{c}{Line luminosity}\\
\cline{4-6} \cline{8-10} \\
\colhead{Num}& \colhead{Y}& \colhead{X}& \colhead{A}& \colhead{B}& 
\colhead{rms (dex)}& \colhead{}& \colhead{C}& \colhead{D}& \colhead{rms (dex)}} 
\startdata
1	&H$\alpha$	&P$\alpha$	&0.074$\pm$0.038	&
0.934$\pm$0.084	&0.045&	&0.916$\pm$0.014&0.961$\pm$0.025&0.141	\\
2	&H$\beta$	&P$\alpha$	&0.105$\pm$0.037	&
1.017$\pm$0.080	&0.057&	&0.444$\pm$0.013&0.910$\pm$0.018&0.188	\\
3	&H$\alpha$	&P$\beta$	&0.076$\pm$0.038	&
0.821$\pm$0.075	&0.051&	&0.985$\pm$0.012&1.008$\pm$0.015&0.117	\\
4	&H$\beta$	&P$\beta$	&0.113$\pm$0.033	&
0.895$\pm$0.068	&0.058&	&0.517$\pm$0.011&0.943$\pm$0.013&0.162	\\
\enddata
\end{deluxetable*}
\clearpage

\begin{figure}
\epsscale{.8}
\plotone{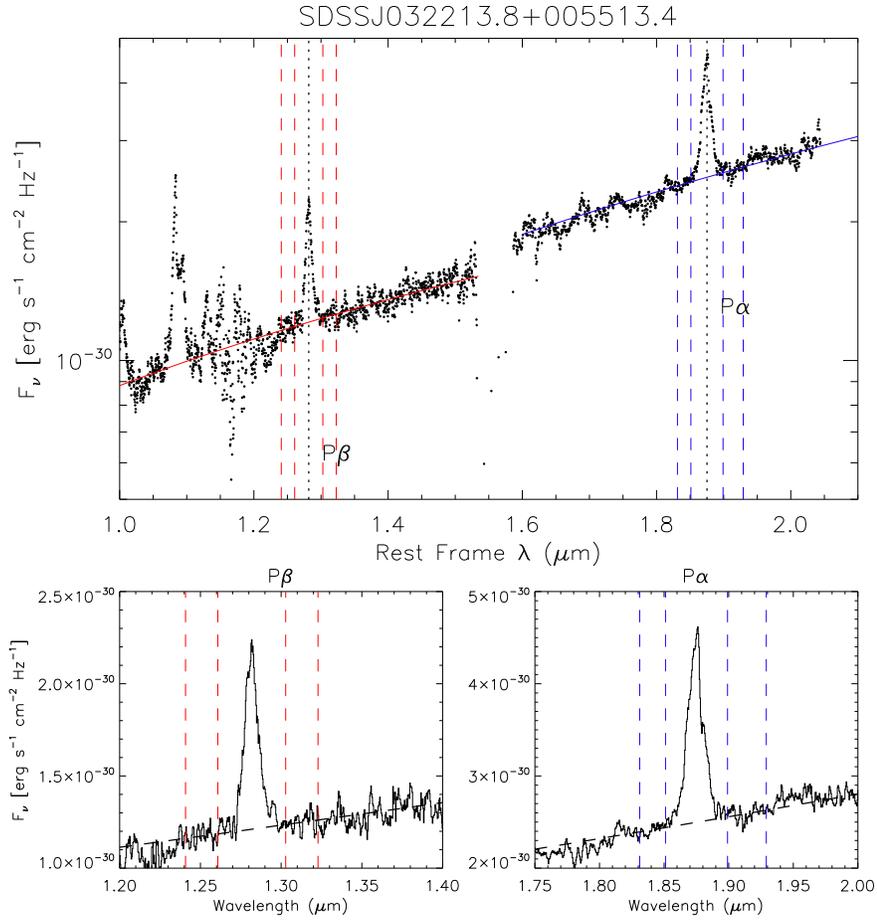}
\caption{An example of the continuum fitting of the G06 NIR spectra.
  On the top panel, we show the NIR spectrum of SDSSJ032213.8$+$005513.4, where
 the blue solid line indicates the continuum fit to the P$\alpha$ wavelength
 region and the red solid line shows the continuum fit to the P$\beta$ wavelength area.
  The wavelength ranges that were used for determining the continuum are indicated
 with the vertical dashed lines.
  In the bottom panels, we show the expanded views around P$\beta$ and P$\alpha$ lines together 
 with the determined continuum.}
\end{figure}

\begin{figure}
\epsscale{.9}
 \includegraphics[scale=0.8]{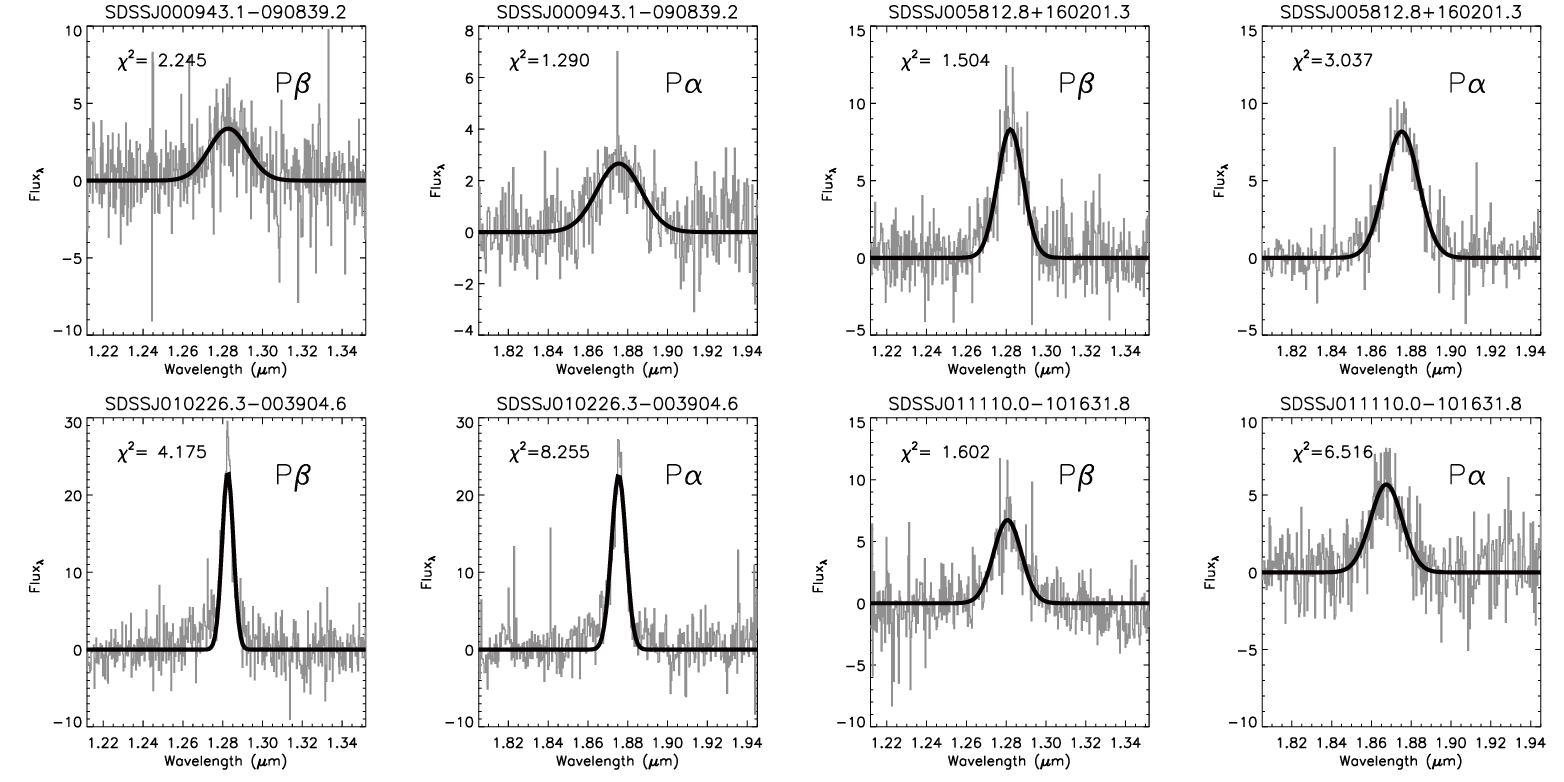}\\
 \includegraphics[scale=0.8]{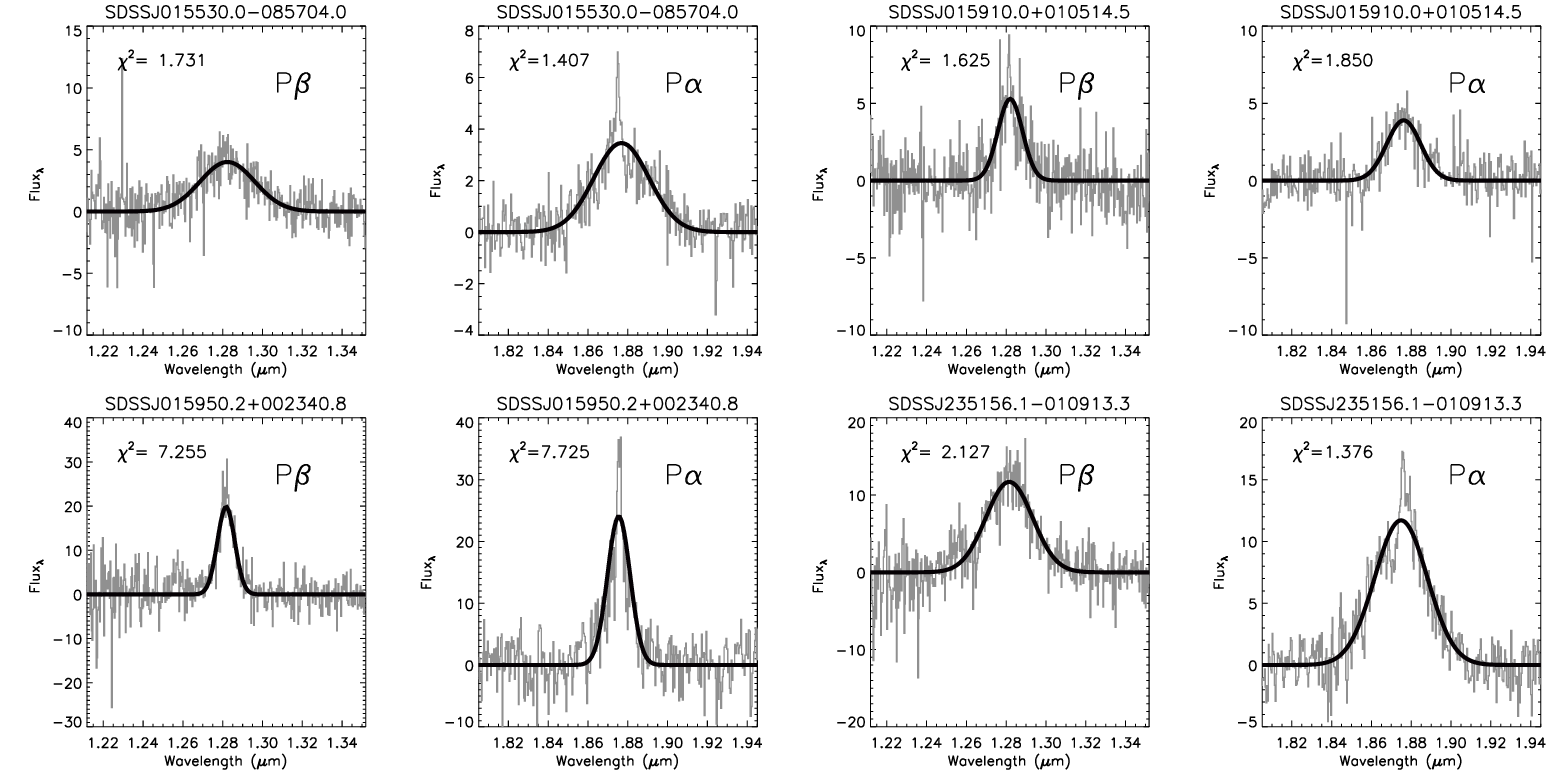}
\caption{Results of the fitting of Paschen lines with a Gaussian
 function to the G06 sample. AGNs with both P$\alpha$ and P$\beta$ measurements
 are shown here. The ordinate is in units of 10$^{-17}~\mathrm{erg~s^{-1}}$ cm$^{-2}~\mu$m$^{-1}$.
 The continuum is already subtracted. The black solid line
 indicates the best-fit and the grey solid line indicate observed spectrum.
}
\end{figure}

\begin{figure}
\epsscale{.9}
\plotone{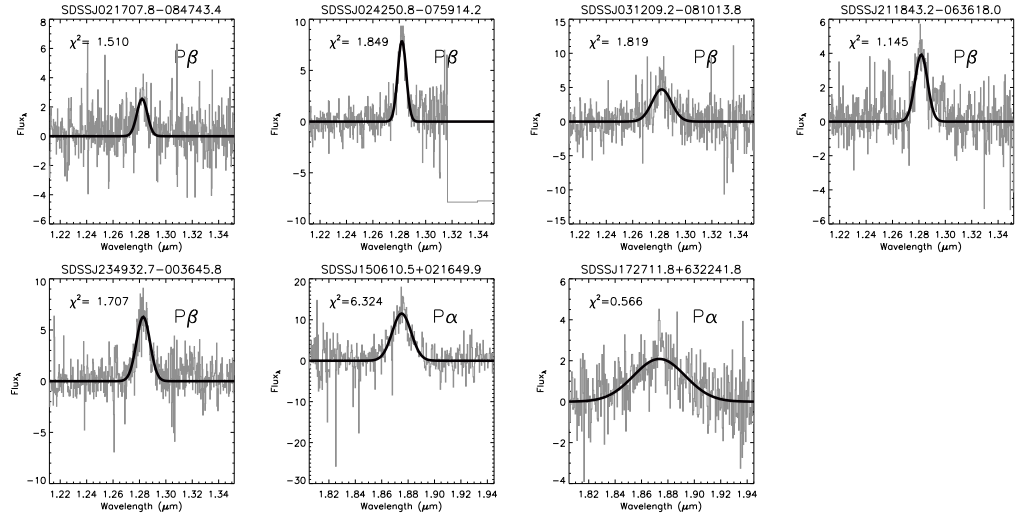}
\caption{Gaussian fits of the
 Paschen lines of the G06 sample. AGNs with only P$\alpha$ or P$\beta$ measurements
 are shown here.  The meaning of the lines is identical to Figure 3.}
\end{figure}

\begin{figure}
\epsscale{0.3}
 \includegraphics[scale=0.5]{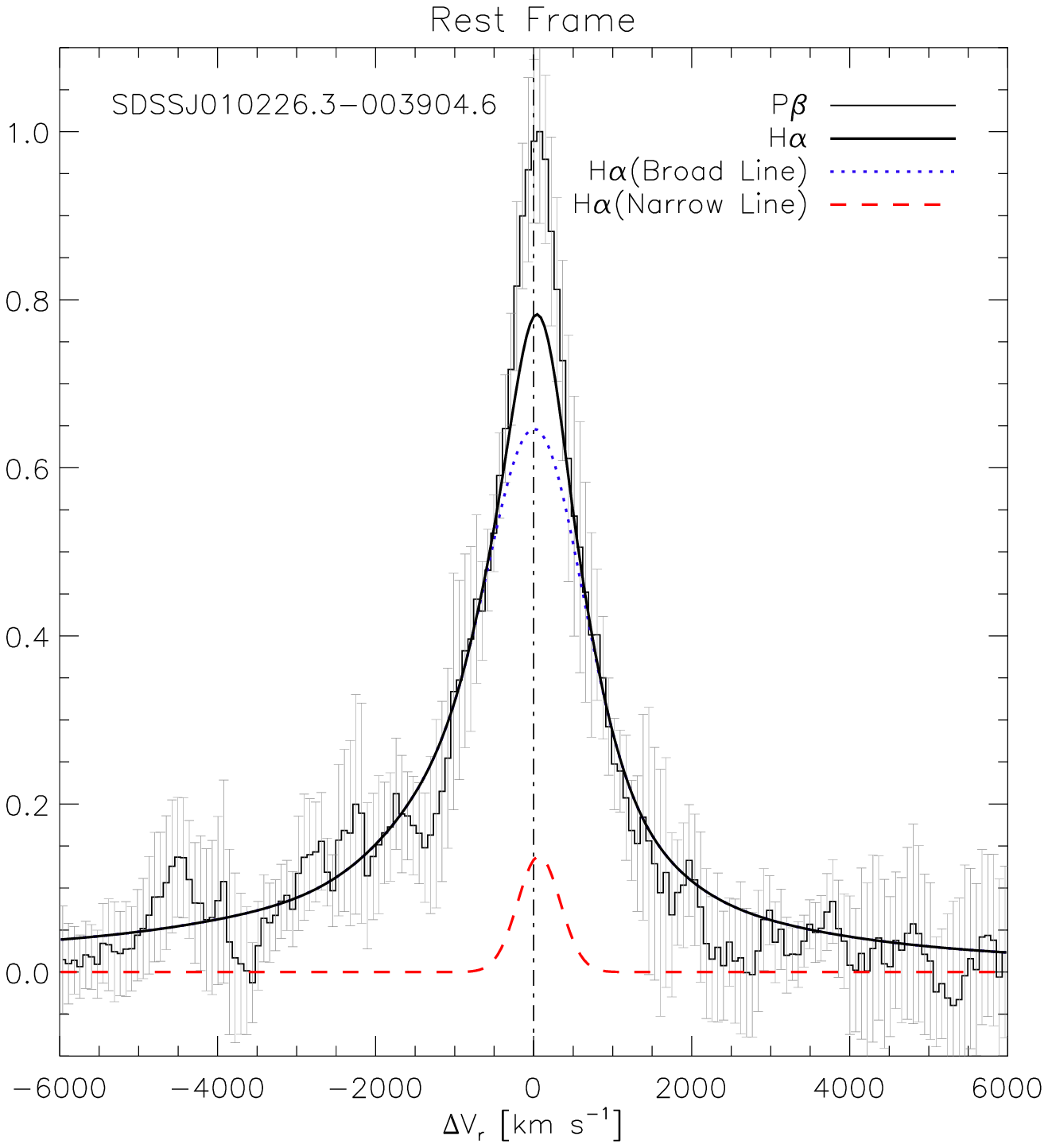}
 \includegraphics[scale=0.5]{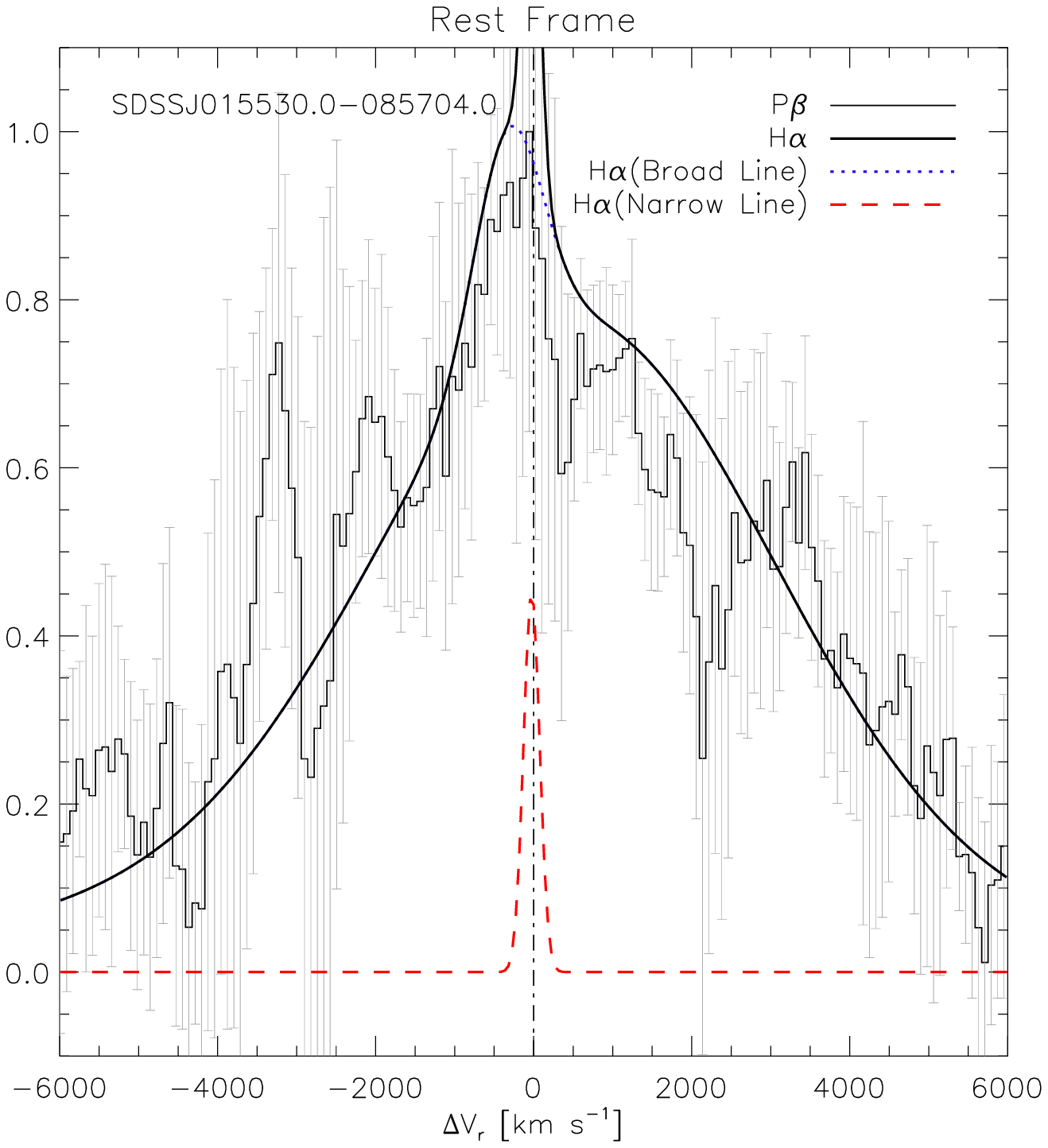}\\
 \includegraphics[scale=0.5]{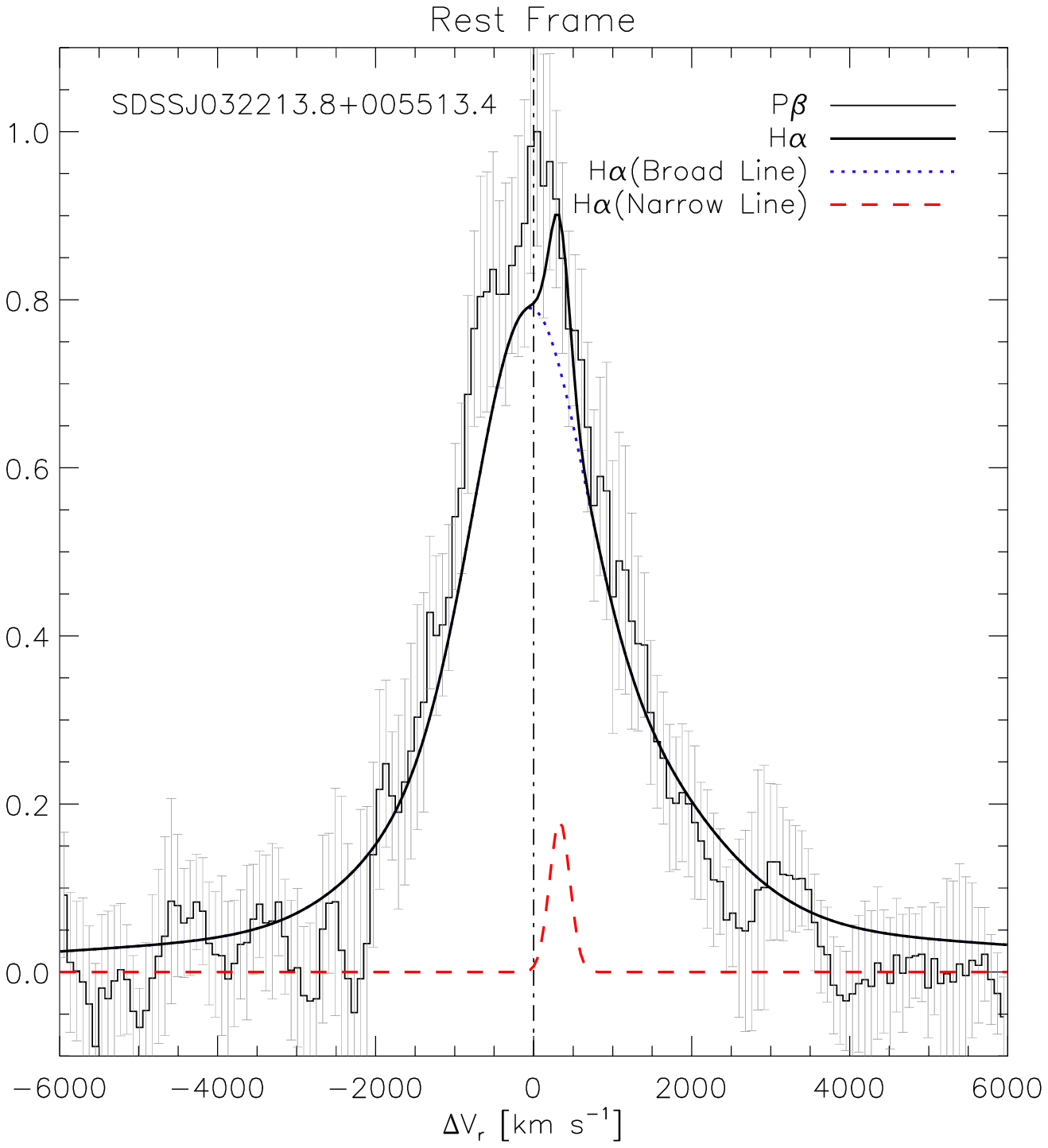}
 \includegraphics[scale=0.5]{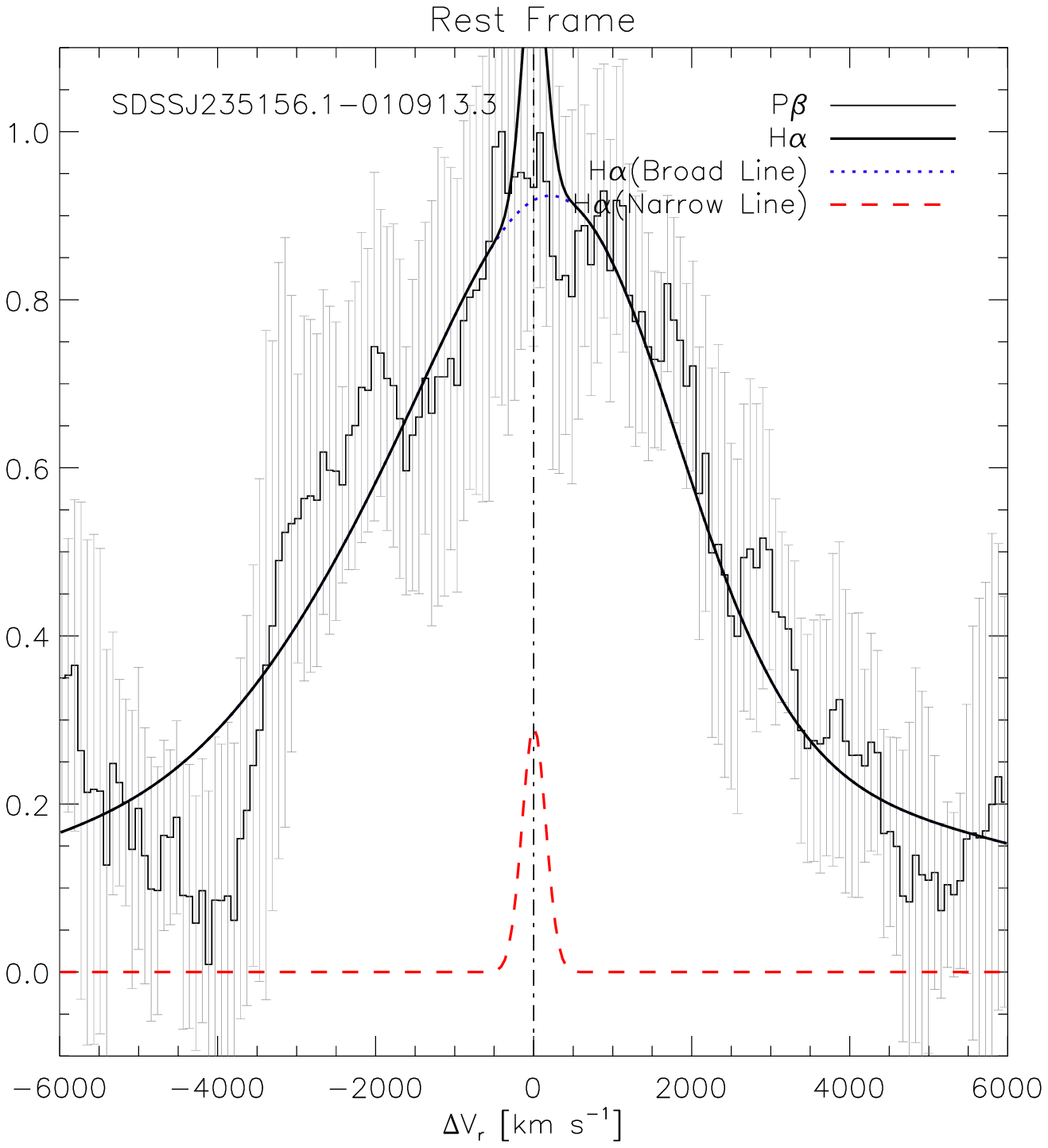}
\caption{The comparison of the P$\beta$ line profiles against the H$\alpha$ line profiles. 
  Only objects with P$\beta$ line S/N $>$ 25 are shown here (histogram with errors).
  The H$\alpha$ lines are indicated with thick solid lines, and 
  the thin dotted lines and the thick dashed lines represent the broad and the narrow components
 of the H$\alpha$ lines. The P$\beta$ profiles are normalized to have the maximum value of 1 
 while the H$\alpha$ are normalized to have the same total flux as the normalized P$\beta$ 
 profiles. The figure shows that the P$\beta$ line profile shapes are consistent with
 the H$\alpha$ profile shapes.}
\end{figure}

\begin{figure}
\epsscale{1.}
\plotone{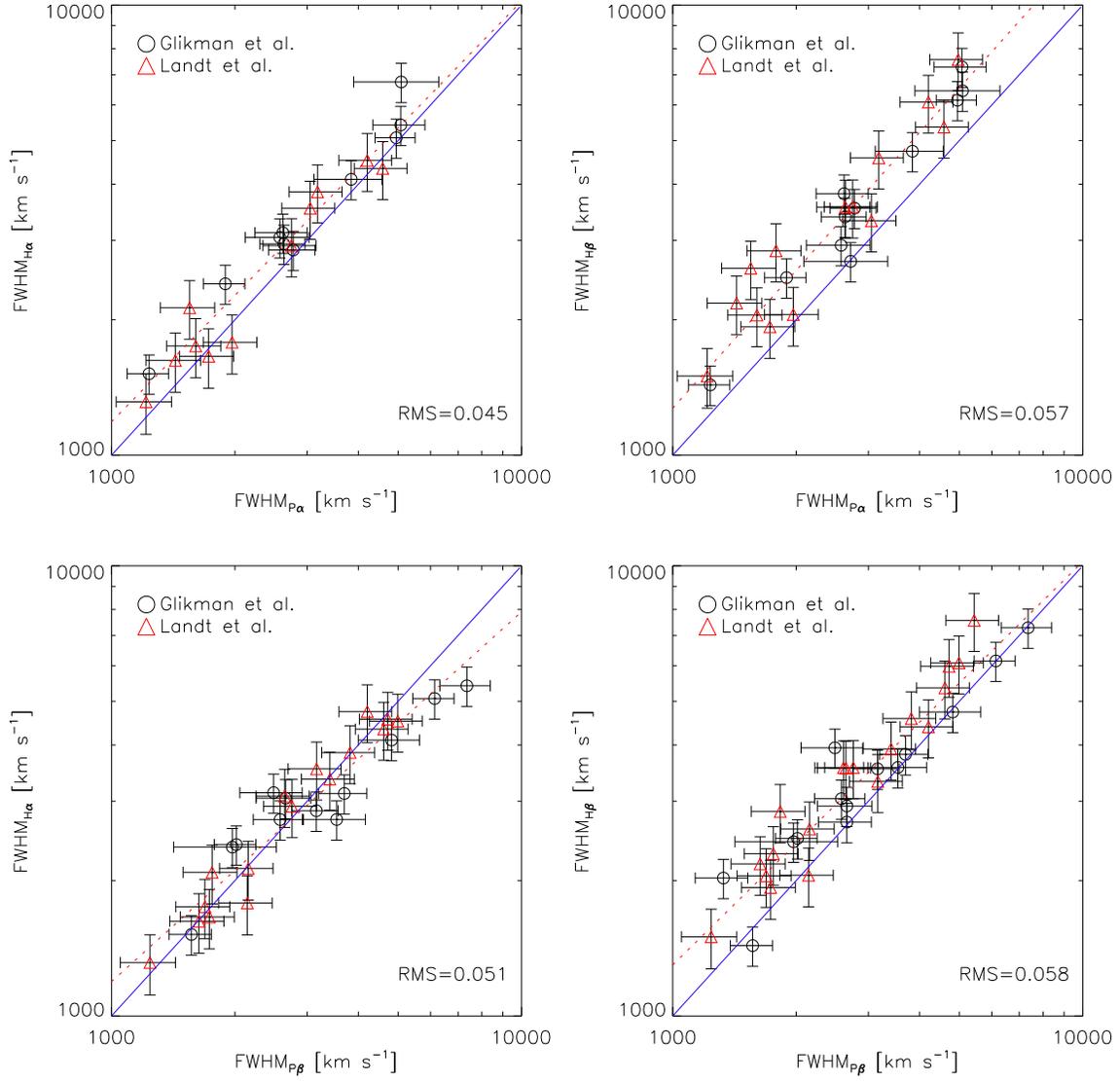}
\caption{The comparison of the FWHM widths of the Paschen lines versus the Balmer
 lines. The top panels compare the P$\alpha$ FWHMs versus the Balmer line FWHMs,
 while the bottom panels show the comparison of the P$\beta$ FWHMs versus the
 Balmer line FWHMs. The open circles are for AGNs from G06, while the red triangles are for the L08
 sample.
 The solid line indicates a line where the Balmer and the Paschen quantities are identical.
 The dotted line indicates the best-fit line between the two quantities as described in the text.
}
\end{figure}

\begin{figure}
\epsscale{1.}
\plotone{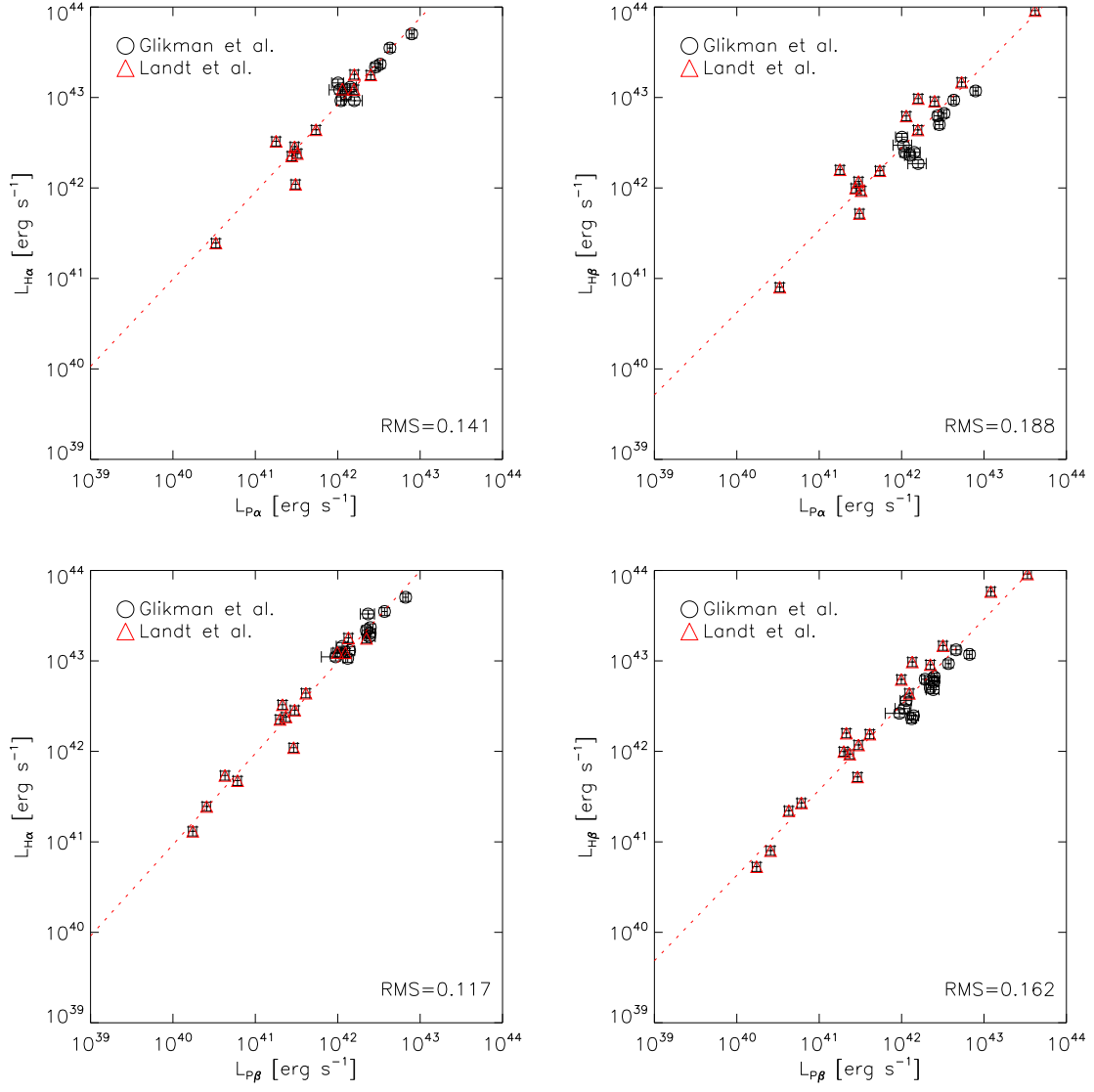}
\caption{The comparison of the Paschen line luminosities against those of the Balmer lines.
 The top two panels compare the P$\alpha$ fluxes against the Balmer lines fluxes,
 while the bottom two panels compare the P$\beta$ fluxes against the Balmer line fluxes.
 The symbols are same as those in Figure 5.
 The meaning of dotted line is identical to that in Figure 5.
}
\end{figure}

\begin{figure}
\epsscale{1.}
\plottwo{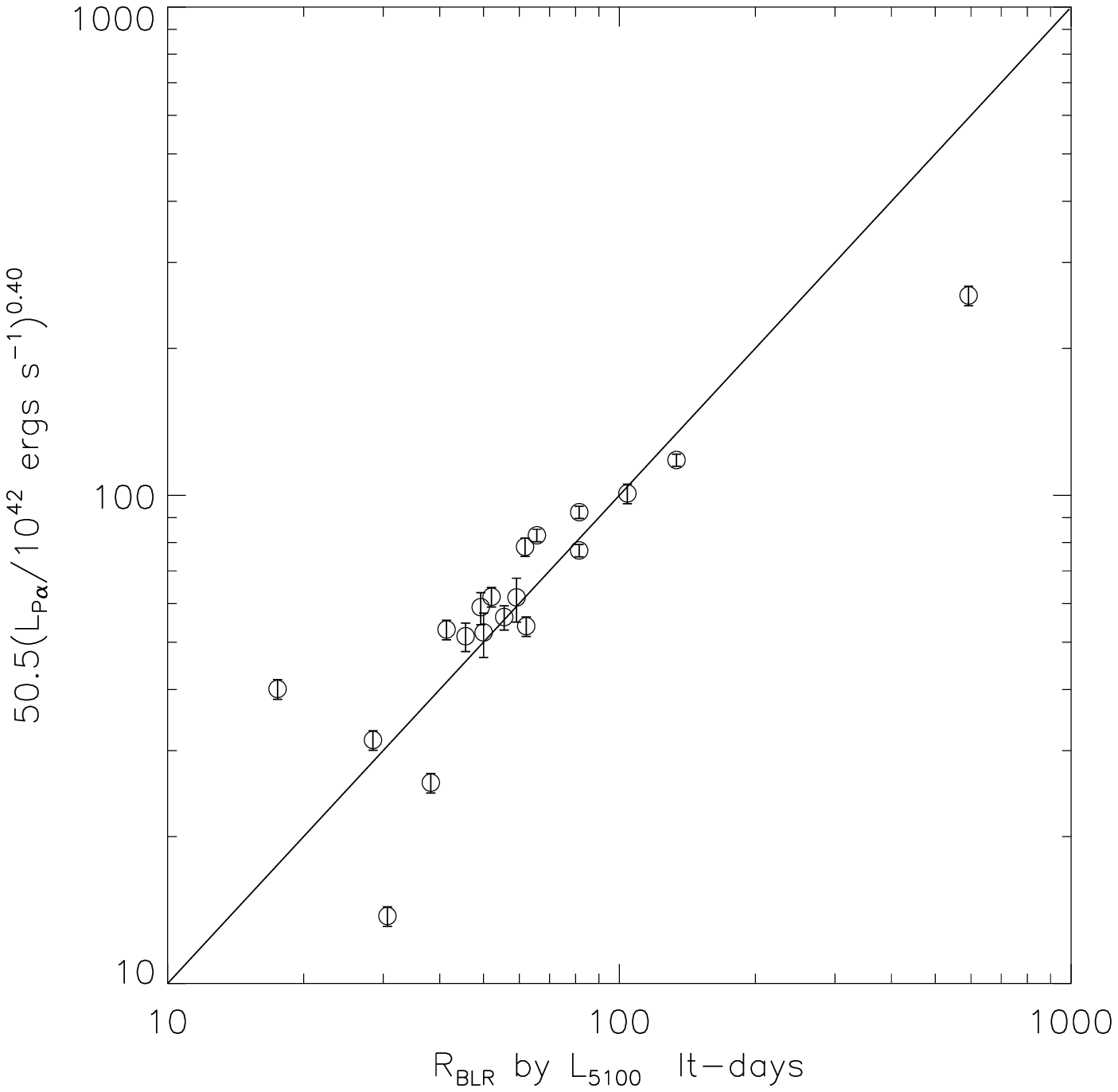}{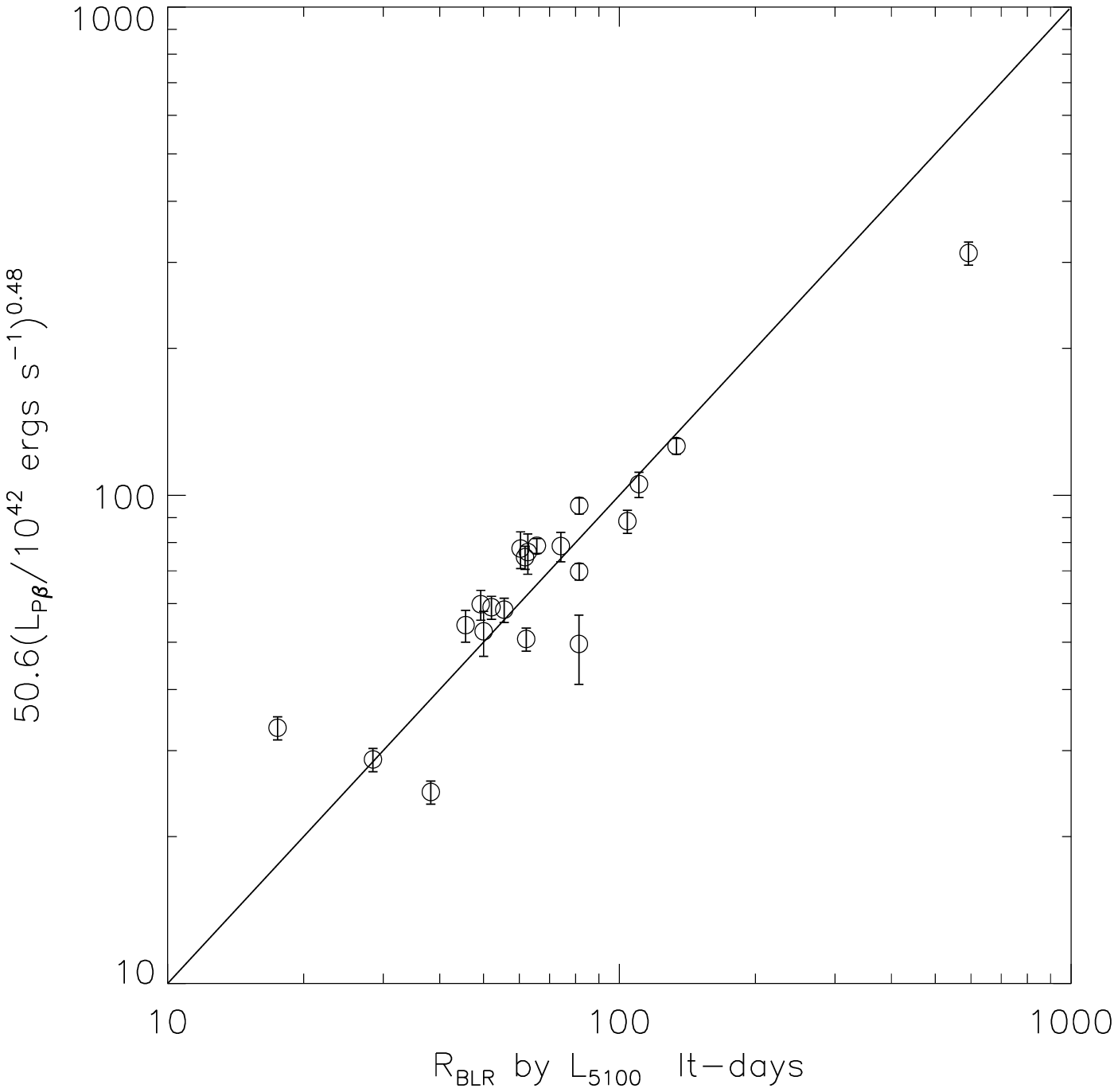}
\caption{The correlation between the P$\alpha$ (left) and the P$\beta$ (right)
 luminosities and the BRL sizes. The BLR sizes are derived from the 5100$\mathrm{\AA{}}$
 luminosities. The good correlation (albeit over 40 lt-days to 110 lt-days)
 suggests that the Paschen line luminosities
 can be used to estimate the BLR size. The solid lines indicate the best-fit relation of
 the correlation.}
\end{figure}

\begin{figure}
\epsscale{1.0}
\plottwo{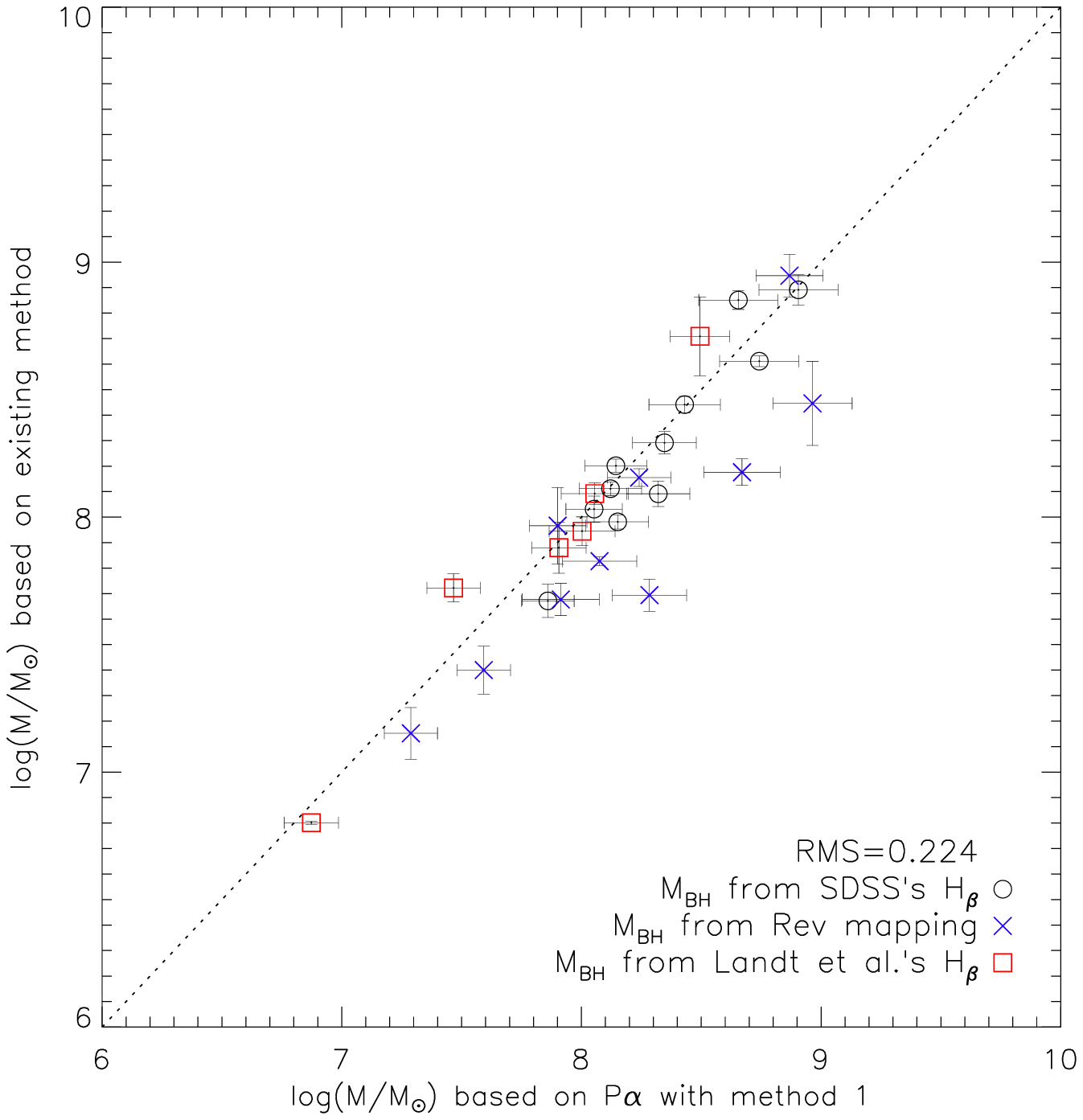}{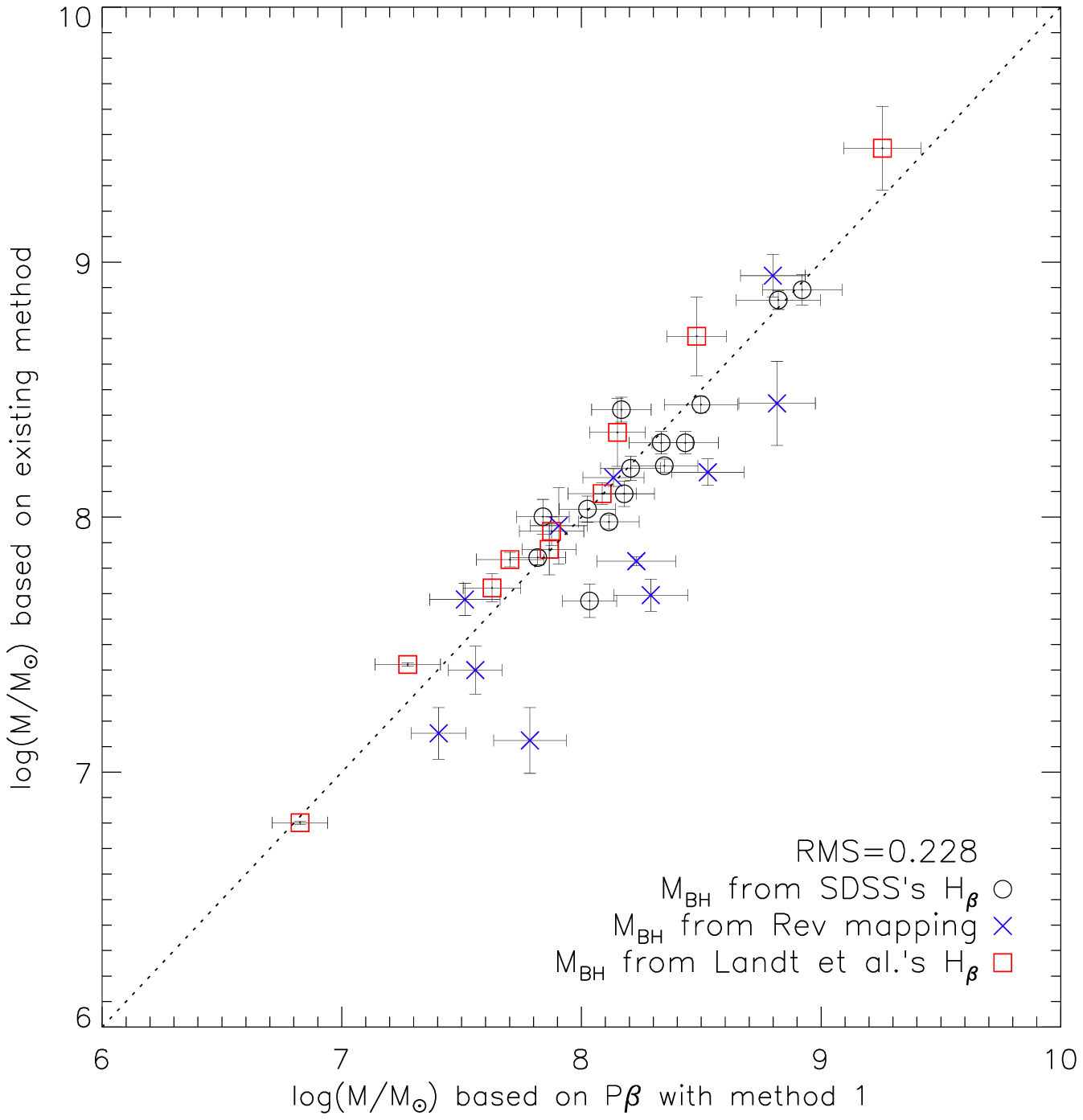}
\caption{Comparison of the Paschen-line based $M_{\rm BH}$ versus 
 $M_{\rm BH}$ derived from the reverberation mapping or optical spectra using
 the 1st method described in the text.
 Black circles indicate the G06 sample. Blue crosses show the L08 sample
 with $M_{\rm BH}$ derived from the reverberation mapping (
 Vestergaard $\&$ Peterson 2006).  Red squares
 are for the L08 sample with $M_{\rm BH}$ estimated from single epoch
 optical spectra.
 The solid line indicates an one to one function with $M_{\rm BH}$ based on existing method and
 it upon properties of Paschen line.
}
\end{figure}

\begin{figure}
\epsscale{1.0}
\plottwo{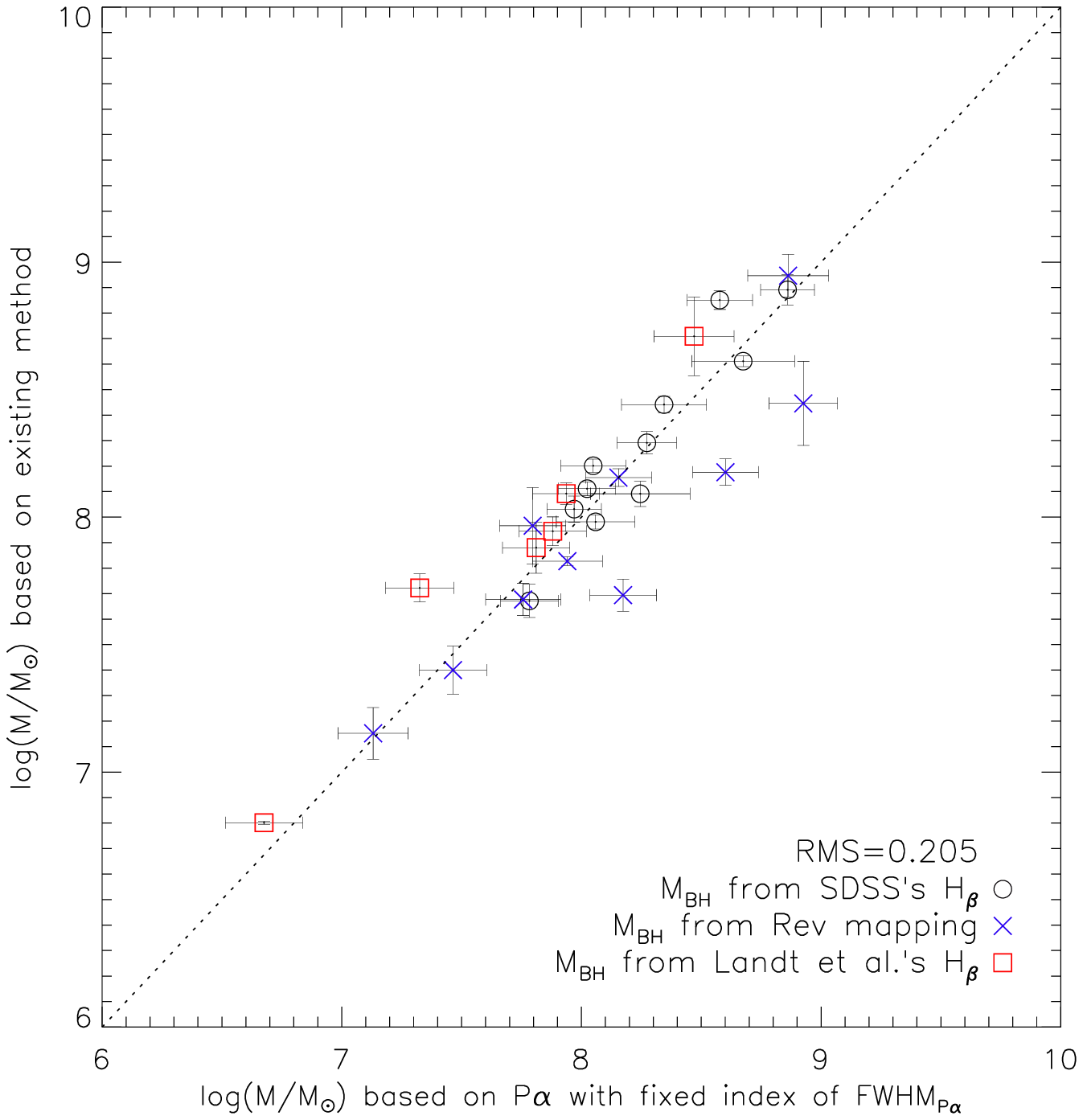}{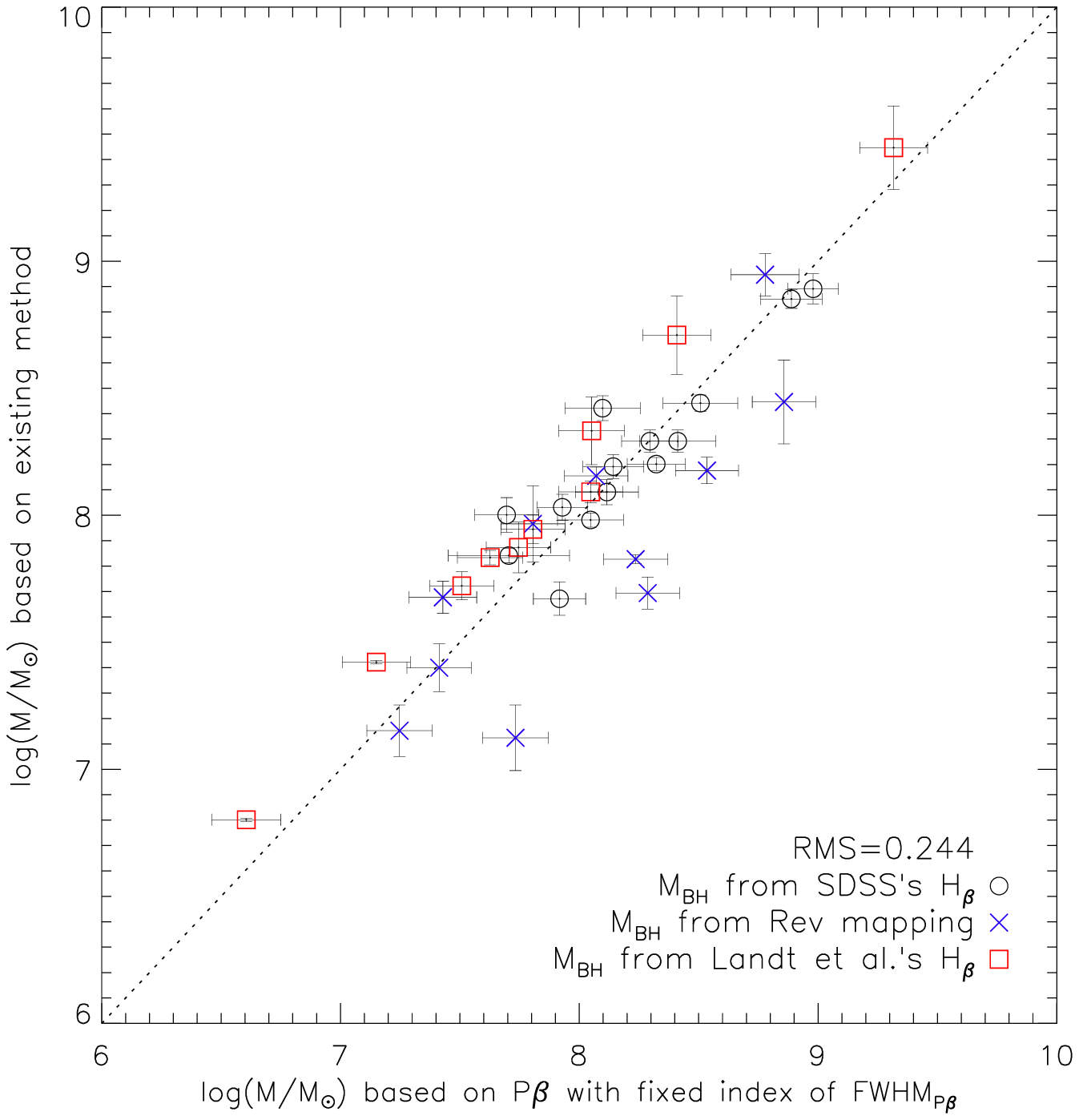}
\caption{
  Similar to Figure 8, but with the Paschen-based $M_{\rm BH}$ where
  $M_{\rm BH}$ are derived with an estimator for which the power-law
  index of FWHM is fixed to 2. The meaning of the symbols
 and line are identical to Figure 8.}
\end{figure}

\begin{figure}
\epsscale{1.0}
\plottwo{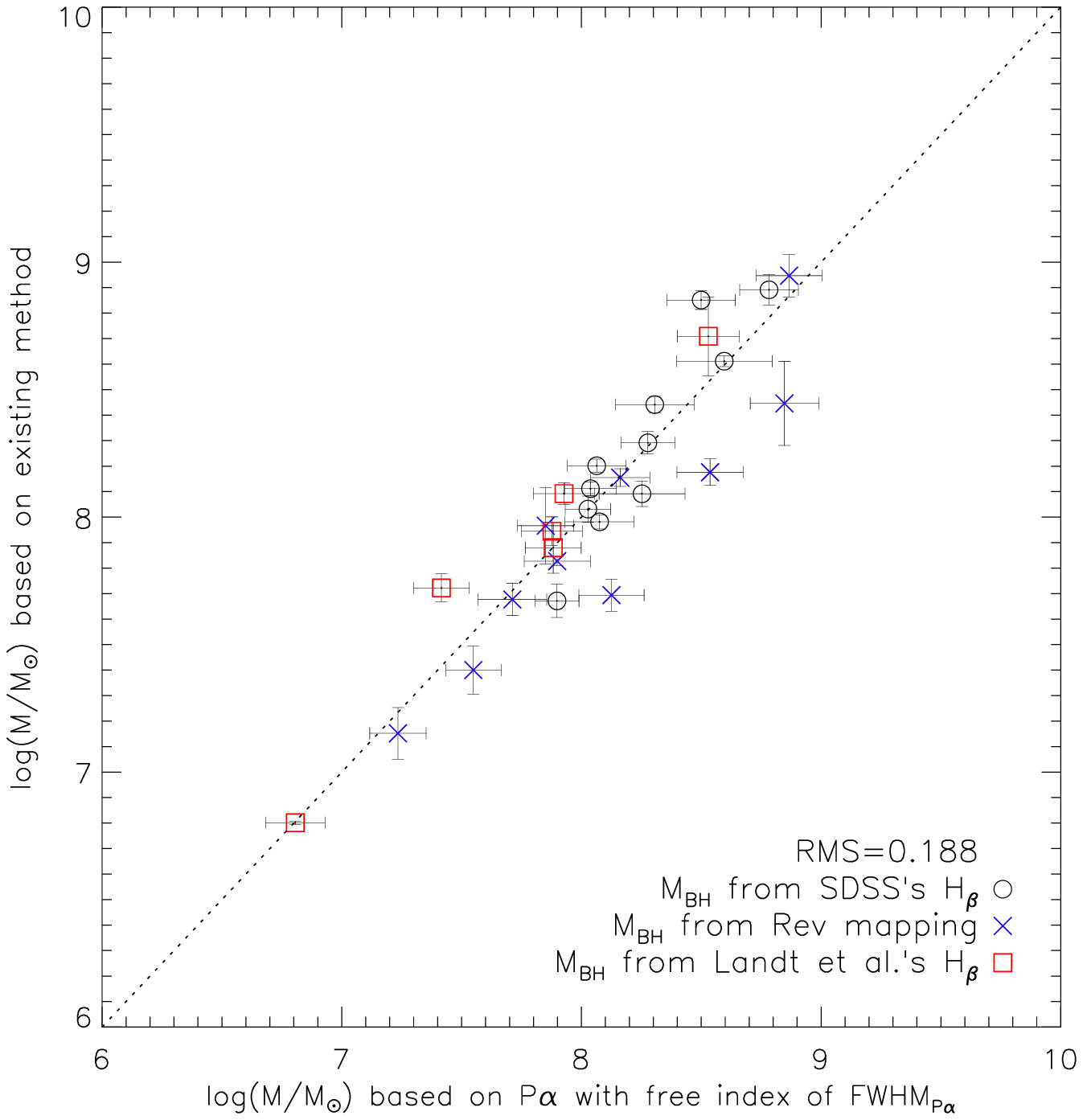}{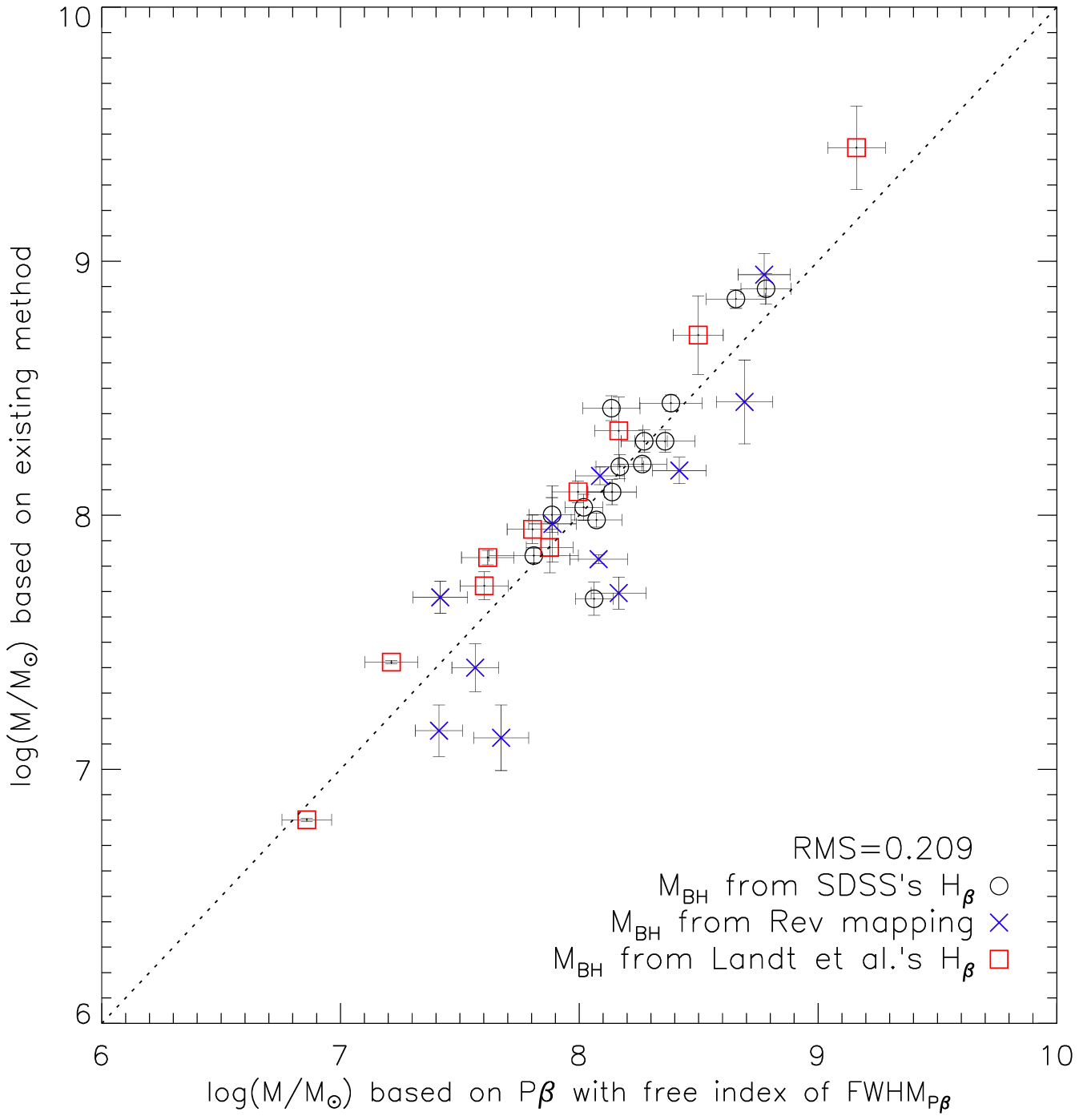}
\caption{
  Similar to Figure 8, but with the Paschen-based $M_{\rm BH}$ where
  $M_{\rm BH}$ are derived with an estimator for which the power-law
  index of FWHM is set free during the fit. The meaning of the symbols
 and line are identical to Figure 8.}
\end{figure}

\begin{figure}
\epsscale{1.0}
\plotone{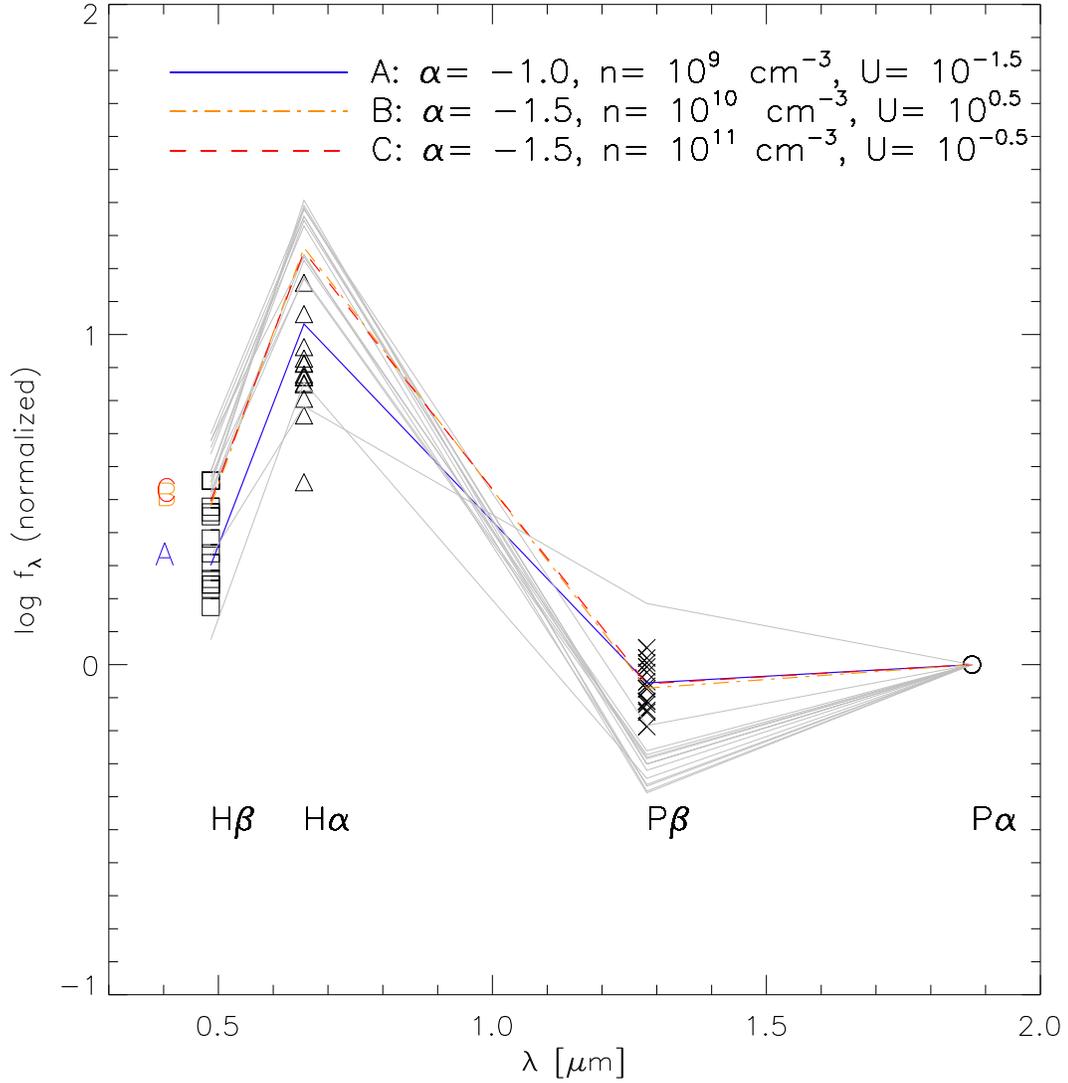}
\caption{Line ratio of type-1 AGNs compared with theoretical models from CLOUDY code.
 Fluxes are normalized to that of P$\alpha$.
 These models show that typical type 1 AGN condition is $\alpha$= -1.0, n= $10^9~\mathrm{cm^{-3}}$
 and U= $10^{-1.5}$. The grey lines are theoretical models from various condition, within
 $\alpha$= -1.5 $\sim$ -1.0, n= $10^{9}~\sim~10^{11}~\mathrm{cm^{-3}}$ and U= $10^{-1.5}~\sim~10^{0.5}$.
}
\end{figure}

\begin{figure}
\epsscale{1.0}
\plottwo{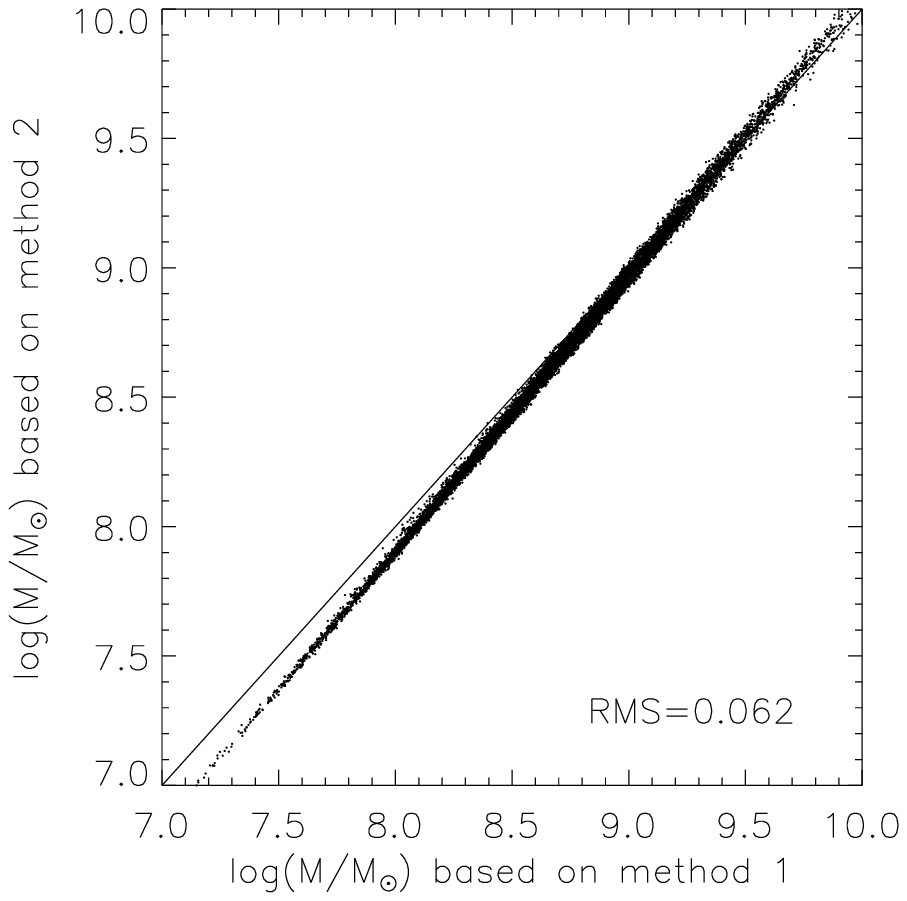}{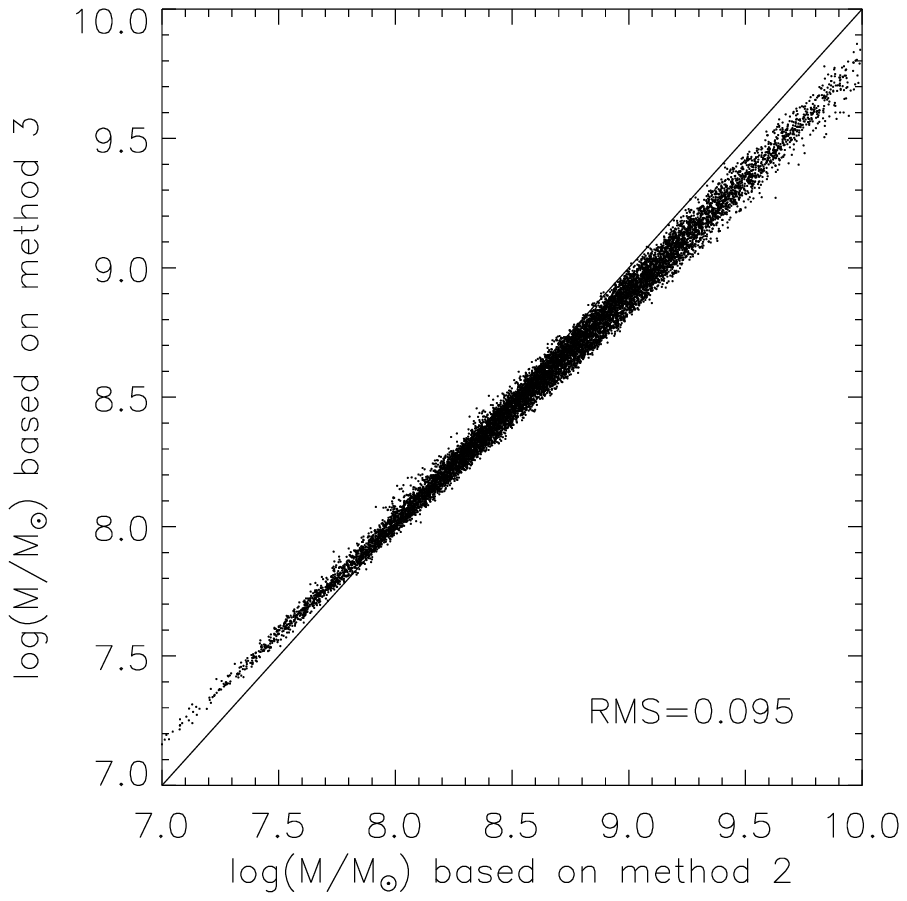}
\caption{The left panel compares $M_{\rm BH}$ from the method 1 and
 the method 2.
  In the right panel, we compare $M_{\rm BH}$ from the method 2 versus the method 3.
  The $M_{\rm BH}$ values agree well with each other in this case. The data points come
 from SDSS DR5 quasars in Shen et al. (2008) where the Paschen FWHM and luminosities 
 are estimated from the corresponding H$\beta$ quantities. The solid line indicates 
 the case where the BH masses derived from different methods are identical.
}
\end{figure}

\end{document}